\documentclass[msom]{informs3} 

\OneAndAHalfSpacedXI

\usepackage{fancyhdr}
\usepackage{natbib}
 \bibpunct[, ]{(}{)}{,}{a}{}{,}%
 \def\bibfont{\small}%

\usepackage{booktabs} 
\usepackage{diagbox}
\usepackage{bbm}
\usepackage{url}
\usepackage{stfloats}
\usepackage{amsfonts}

\usepackage{xcolor}
\usepackage{epstopdf}
\usepackage{algpseudocode}
\usepackage{bm}
\usepackage{multirow}

\usepackage[labelformat=simple]{subcaption}
\usepackage{adjustbox}
\usepackage{algorithm}

\usepackage{caption}

\newtheorem{observation}{Observation}

\TheoremsNumberedThrough     

\EquationsNumberedThrough    

\begin{document}




\TITLE{The Impact of Autonomous Vehicles on Ride-Hailing Platforms with Strategic Human Drivers 
\thanks{This is a working paper. Shuqin Gao and Xinyuan Wu should be considered as co-first authors.}
}

\ARTICLEAUTHORS{%
\AUTHOR{Shuqin Gao}
\AFF{The Chinese University of Hong Kong, Shenzhen, China, 
\EMAIL{gaoshuqin@cuhk.edu.cn}}
\AUTHOR{Xinyuan Wu}
\AFF{The Chinese University of Hong Kong, Shenzhen, China, \EMAIL{xinyuanwu@link.cuhk.edu.cn}}
\AUTHOR{Antonis Dimakis}
\AFF{Athens University of Economics and Business, Athens, Greece, \EMAIL{dimakis@aueb.gr}}
\AUTHOR{Costas Courcoubetis}
\AFF{The Chinese University of Hong Kong, Shenzhen, China, \EMAIL{costas@cuhk.edu.cn}}

} 

\ABSTRACT{
Motivated by the rapid development of autonomous vehicle technology, this work focuses on the challenges of introducing them in ride-hailing platforms with conventional strategic human drivers. We consider a ride-hailing platform that operates a mixed fleet of autonomous vehicles (AVs) and conventional vehicles (CVs), where AVs are fully controlled by the platform and CVs are operated by self-interested human drivers. Each vehicle is modelled as a Markov Decision Process that maximizes long-run average reward by choosing its repositioning actions. The behavior of the CVs corresponds to a large game where agents interact through resource constraints that result in queuing delays. In our fluid model, drivers may wait in queues in the different regions when the supply of drivers tends to exceed the service demand by customers. Our primary objective is to optimize the mixed AV-CV system so that the total profit of the platform generated by AVs and CVs is maximized. To achieve that, we formulate this problem as a bi-level optimization problem $\mathcal{OPT}$ where the platform moves first by controlling the actions of the AVs and the demand revealed to CVs, and then the CVs react to the revealed demand by forming an equilibrium that can be characterized by the solution of a convex optimization problem. We prove several interesting structural properties of the optimal solution and analyze simple heuristics such as AV-first where we solve for the optimal dispatch of AVs without taking into account the subsequent reaction of the CVs. We propose three numerical algorithms to solve $\mathcal{OPT}$ which is a non-convex problem in the platform decision parameters. We evaluate their performance and use them to show some interesting trends in the optimal AV-CV fleet dimensioning when supply is exogenous and endogenous. 
}



\maketitle


%


\section{Introduction}


As the technology of autonomous vehicles (AVs) becomes more mature with decades of development, there is a supported belief that AVs will be widely deployed by ride-hailing platforms to significantly reduce the cost of ride-hailing services \citealt{tata}. Leading platforms, like Uber, Lyft, and Didi have made significant investments in the development of AV technology and the related research \cite{uber, lyft, didi}. In the foreseeable future, these platforms are expected to commence operations with a mixed fleet of AVs and Conventional Vehicles (CVs) driven by humans, at least in the initial phase. Significant research \citealt{av,Wigand,Siddiq,Ryzin} has been dedicated to understanding the impact of AV operation on the platform, the riders, and the existing human drivers in the ride-hailing system. 

A key problem for the platform is how to operate the mixed fleet of AVs and CVs over a spatial network that corresponds to a given city. Since the AVs are under the direct ownership and control of the platform, it is rather straightforward for the platform operator to design the optimal empty-car repositioning/routing strategy for AVs to serve the largest amount of demand across regions and maximize profit. However, differing from the fully-controlled AVs, human drivers are self-interested and make their own operational decisions strategically, including when and where to reposition, as well as their initial decision to work for the platform \citealt{purecv}. This difference in control structure for the mixed fleet vehicle types creates a practical challenge to the platform: design an optimal strategy for the AVs, together with an appropriate demand allocated to CVs to mitigate the inefficiencies resulting from the selfish behaviors of CVs. 

A key component in the solution of the mixed fleet optimization problem is the prediction of the behavior of the CVs. There is a continuous interaction between strategic human drivers that suggests the concept of an equilibrium. This interaction is in terms of the waiting times to get assigned to customers in each of the different regions. When many drivers prefer to serve customers in the same region leading to more supply than demand, there will be longer waiting times to get an available customer. This, similar to repositioning (driving empty to some other region besides the one they deposited a customer), affects the average reward per unit time from trips, since it implies a longer average time to collect a unit of reward. Each individual driver is too small to affect average waiting times in the system and optimizes its actions assuming these times are fixed. However, the collective behavior of drivers will have an effect, suggesting a game-theoretic model. 

Assuming that the demand remains stationary for a long enough time, the CV system will reach an equilibrium. Since repositioning times might be larger than waiting times, self-interested CV drivers may strategically queue up in high-profitable regions instead of repositioning to other regions and serving more demand, implying social inefficiency and reduced platform revenue. With the introduction of AVs, the platform can mitigate such a loss by allocating to AVs some of the demand in the high-profitable regions where CVs prefer to wait. Then, by reducing the available demand for CVs, CV waiting times will be increased and CVs will be incentivized to reposition and serve more demand in remote regions. Alternatively, the platform could choose not to assign the high-profitable demand to AVs and let them serve these remote demand even though AVs always generate a higher profit to the platform than CVs, which might be counter-intuitively better. This shows the complexity and the interrelation in the management of AVs and CVs.

A crucial component to make the above analysis possible is the ability to calculate the CV equilibrium for a given demand and region topology. To the best of our knowledge, this was not possible up to now except for the special case of two-region networks \citealt{purecv,av}, and the analysis does not extend to a larger number of regions. In this work, we analyze for the first time CV equilibria for an arbitrary number of regions. This allows us to precisely define and solve the mixed AV-CV fleet optimization problem for general networks, and prove certain important properties of the optimal solution.

\textbf{Results and Contributions.} To study the mixed-fleet AV-CV optimization problem, we begin with the base case scenario where all the vehicles are CVs and the demand revealed by the platform to CVs is assumed given. We model the collective behavior of the CVs as a mean-field game of the type analyzed in \citealt{antonis}, where a continuum of agents, each solving an MDP, interact in a fluid model by queuing for access to resources that get replenished at a fixed rate. In our case, we correspond the states of the MDP to the geographical regions, actions to repositioning decisions, and resources to customer arrivals in each region. Using the mean-field game equilibrium analysis in \citealt{antonis}, the CV equilibrium can be computed by solving a convex optimization problem. Further, we prove the uniqueness of the platform profit generated by CVs in the possible equilibria and its increasing monotonicity with respect to CV fleet size.

Next, we address the optimum management of the mixed fleet case. Based on the program for obtaining the CV equilibrium for arbitrary customer demand, we define the bi-level optimization problem $\mathcal{OPT}$ to maximize the total profit generated by CVs and AVs. At the upper level, the platform (as the leader) decides on i) the demand revealed to CVs and ii) the state-action rates of AVs to serve the residual demand. Then, at the lower level, CVs (as the follower) will respond by forming an equilibrium for the assigned demand as determined by the convex problem mentioned above. Note that in our formulation of the bi-level problem, there is a subtle difference with earlier proposals\footnote{Derived so far only for the special case of a two-region network \citealt{purecv,av}.}: we don't decide first on the amount of demand served by AVs and then let CVs serve the residual demand. This is not necessarily optimal, since in order to maximize profit we may need to restrict even further the demand served by CVs. This is because the profit generated by CVs is not monotone in their allocated demand, and we might do better by hiding some demand from CVs even if this is not served by the AVs. The implication of this is that $\mathcal{OPT}$ is not a convex problem and finding its global optimum can be challenging.

Following the definition of $\mathcal{OPT}$, we prove certain important properties of the solution and propose three computational procedures to numerically solve it. More specifically, we obtain the following results.



\begin{itemize}

    \item We prove that the profit of the platform is non-decreasing in the numbers of CVs and AVs.
    
    \item We prove that in the optimal solution, if there are any CVs serving the residual demand, then all the AVs must be fully utilized. 
    
    \item One might thus argue that the `AV-first' policy is optimal: the platform should deploy the available AVs to maximize its AV revenue without considering the existence of CVs. We show that this is not necessarily optimal. We give an example showing that the platform may prefer to offer to CVs demand in regions requiring low repositioning costs and use AVs to serve other regions with higher such costs that otherwise would not be served by CVs under AV-first.
    
    \item Motivated by this, we further prove the AV-first is optimal if: i) there exists an AV strategy that can serve all demand, or ii) under the AV-first, the residual demand 
    is fully served by the CVs in the resulting equilibrium. 
    
    \item Given the simplicity of AV-first, we like to investigate its performance loss compared to $\mathcal{OPT}$.  We rigorously prove that in any two-region network, the performance loss of the AV-first policy is $20\%$ at most. A comprehensive numerical analysis for networks with a larger number of regions motivates us to conjecture that this $20\%$ upper bound holds in general.

    \item We propose three algorithms to solve $\mathcal{OPT}$ for general networks: i) a gradient-descent algorithm that searches the local maximum in the direction of the gradient; ii) a bundle method algorithm designed for addressing non-smooth optimization problems, leveraging subgradients from preceding iterations; and iii) a genetic algorithm that adaptively searches the solutions inspired by the evolutionary process of natural selection. We show that all of our proposed algorithms attain the optimum in the fundamental two-region networks, where we can derive analytically the optimal solution. Further, we conduct extensive numerical simulations for a broader scope of grid networks, showing the performance improvement of our proposed algorithms as compared to the simple AV-first policy baseline.

    \item Finally, we extend our analysis to the endogenous supplies of AVs and CVs when the platform can decide the amount of AVs to purchase and CV drivers can decide whether to work for the platform. Our proposed algorithms are suitable for solving this extended problem with slight modifications. Numerical results demonstrate the efficiency of the modified algorithm in the endogenous scenario.

\end{itemize}

\textbf{Related Literature.} The existing body of literature related to our work can be categorized into two primary streams: i) the optimization in the traditional ride-hailing platforms with only CVs, addressing issues like repositioning, matching and pricing, and ii) the integration of AVs into ride-hailing systems. 

In the first stream analyzing pure CV systems, research on empty-car repositioning over a spatial network is exemplified by works such as \citealt{dai,Benjaafar2022,purecv}. Of particular relevance to our study is \citealt{purecv} which uses a steady-state fluid model, as we do, to analyze the behavior of strategic CV drivers, but limiting to a two-region network and not considering the AVs. \citealt{dai} also uses a fluid model, but studies the repositioning strategy for vehicles that are centrally controlled. They prove the optimal network utility obtained from the fluid-based optimization is an upper bound on the utility in the practical finite car system for any routing policy. \citealt{Benjaafar2022} models the system by utilizing a closed queuing network, where demand arises following a Poisson process. They obtain explicit and closed form lower and upper bounds on the minimum number of vehicles needed to meet a target service level. Moreover, there is a more recent work \citealt{Candogan} that studies how the platform can utilize information about the underlying state to influence resource repositioning decisions and ultimately increase commission revenues in the broader area of spatial resource allocation.

A larger size of this stream is dedicated to studying online and offline dispatching policies. \citealt{Bertsimas} devise online matching algorithms aimed at maximizing overall profit by efficiently serving customers within specified pickup time windows. From historical simulations using real New York City data\cite{nyc}, they show that their algorithms can dispatch in real-time thousands of taxis. \citealt{Ward} delves into matching customer requests with nearby drivers in the context of time-varying demands and times willing to wait. They use a fluid model approximation and prove the asymptotic optimality of a static matching in a large market regime. \citealt{Banerjee2} propose a family of simple state-dependent matching policies to dynamically manage the geographical distribution of supply, aiming to minimize the proportion of dropped requests in steady state. There is also a branch of works on pricing design in ride-hailing platforms with dynamics \citealt{Benjaafar,Banerjee1,costas-infocom,Chen,Balseiro}), but this is not in the scope of this work.

In the second stream that considers the integration of AVs into ride-hailing systems, most papers focus on studying whether the introduction of automation is harmful or beneficial to platforms, human drivers, and customers \citealt{Wigand,Ryzin,Siddiq}. \citealt{Siddiq} employs a game-theoretic model to characterize the effect of platforms’ access to AVs on platform profit, driver welfare, and social welfare. \citealt{Ryzin} develops an economic model to study the price impact of AVs in different demand scenarios. However, these works overlook the inherent spatial features of vehicle resources and customer demand in ride-hailing systems. The most related work to ours is \citealt{av}, where they also consider a platform operating a mixed fleet of fully-controlled AVs and strategic CVs and study the spatial repositioning of vehicles. Many important questions related to the mixed AV-CV problem have been formulated for the first time in \citealt{av}, but their study is restricted to an asymmetric two-region network and lacks of a scalable solution for the general case of multiple regions. Another interesting topic in this stream is the regulatory policies to mitigate the negative impacts of AVs on ride-hailing systems and improve transport equity \citealt{Gao}.

Another area related to our work is on mean-field games that study the repeated interactions between a large number of strategic agents \citealt{Lasry,Huang}. The characterization of the CV equilibrium in this present paper is an application of the results in \citealt{antonis} that studies the equilibrium of general stationary mean field games in finite spaces with queuing interactions. Based on this characterization, we are able to define and solve the mixed AV-CV optimization problem for general networks. An interesting way to understand our model is to think of the case where agents are not strategic. In this case, our model becomes a closed network of queues. Agents belong to different classes in this closed network and interact by queuing in a finite set of resources.  Kelly in \citealt{Kelly} has considered the steady state performance and the sensitivity analysis when the populations of the different agent types are fixed. In \citealt{antonis} and in this work, we study the equilibrium for the case where agents are strategic and choose to change their type according to the average reward obtained by belonging to the different classes. Agent classes correspond to policy cycles defined later in the paper.

\textbf{Plan for the Paper.} In Section 2, we describe our Markov Decision Process model. In Section 3, we characterize the equilibrium of selfish drivers of CVs. In Section 4, we characterize the mixed-fleet equilibrium including the behaviors of both the platform and the selfish drivers, and, in Section 5, we analyze two interesting properties of the mixed-fleet equilibrium. In Section 6, we propose three algorithms to solve the mixed-fleet equilibrium. In Section 7, we compare the performance and running times of the three algorithms. In Section 8, we generalize our model to endogenous supplies of AVs and CVs and extend the algorithms to solve the endogenous model. In Section 9, we offer conclusions. Proof for all the results are included in the Appendices.

%


\section{System Model}

Consider a ride-hailing platform that operates a mixed fleet of autonomous vehicles (AVs) and conventional vehicles (CVs), where AVs are controlled by the platform and CVs are operated by selfish human drivers. Let $M$ and $N$ denote the fleet sizes of AVs and CVs, respectively\footnote{In the later Section \ref{sec:endo}, we extend our study to the endogenous supply of AVs and CVs under different AVs' purchase costs and CV drivers' opportunity costs when the platform can decide the amount of AVs to purchase and CV drivers can decide whether to work for the platform.}. For each vehicle, upon completing a trip transporting a customer, it becomes `empty', and the action it takes can be categorized into two types: i) stay at the current region to wait for its next customer, or ii) reposition to another region and stay there to get its next customer. The actions of AVs are controlled by the platform while the actions of CVs are selected by human drivers to maximize their own expected earnings.

In this paper, we model the mixed-fleet ride-hailing system operating in a geographical area that is divided into $L$ regions. Let $b_{ij}$ denote the demand rate of customers who originate in region $i$ and have region $j$ as the destination. Then, the total demand in region $i$ is $b_i = \sum_{j=1}^L b_{ij}$, and the customer routing probability is $q_{ij} = b_{ij}/b_i$, i.e., the proportion of customers in region $i$ requesting to transport to region $j$. We use $\bm{b}=\{b_i\}$ to denote the demand vector consisting of demand in each region.

We denote the travel time from region $i$ to region $j$ by $t_{ij}$, where $t_{ij}>0$ if $i\neq j$ and $t_{ii}=0$. In this work, we assume $t_{ij}$ is defined exogenously and is constant, ignoring the effect of platform vehicle service on road congestion and transportation delays.
The platform charges customers a price per unit of distance traveled and a constant average vehicle speed is assumed. This translates to a fixed price $p$ per unit of travel time and thus the received payment for transporting a customer from region $i$ to region $j$ is $pt_{ij}$.

To analyze the action strategy of AVs and CVs, we model each vehicle as a Markov Decision Process (MDP). We assume both types of vehicles share the same state space $\mathcal{S} = \{1,2,..., L\}$ and action space $\mathcal{A}= \{1,2,..., L\}$. We define the state of a vehicle as the region it drops off a customer. The action $\alpha$ of a vehicle in state $i$, i.e., in the region it dropped a customer, corresponds to the choice of the next region it will serve a customer: $\alpha=i$ if it chooses to stay in the same region $i$ to serve customers or $\alpha=j$ if it chooses to reposition and drive empty to region $j\neq i$ to serve customers there.

Under action $\alpha$, a vehicle remains in state $i$ until it finishes serving the next customer from region $\alpha$, and transitions to a new state $k$ with probability $q_{\alpha k}$ (i.e., the probability a customer in the region $\alpha$ goes to region $k$)\footnote{There are many equivalent ways to model actions. For example, we could have in each region the action to stay at that region to wait for its next customer or to reposition to some other region without necessarily committing to serve a customer there. The reader can convince herself that any distribution over this type of actions can be translated into a distribution over actions defined in our model where the driver always picks regions to serve customers.}. A (stationary) strategy consists of choosing in each state a distribution over the set of possible actions $A$.

We use a fluid model to describe the interaction of customers and serving vehicles where vehicles can wait but customers don’t wait. We assume a customer is lost if there is no available vehicle in the region upon its arrival, but vehicles will wait and queues will form if the supply rate of vehicles serving a region tends to exceed the demand rate of arriving customers in the region. Thus, for a vehicle taking action $\alpha$ in state $i$, the expected action execution time $\tau_{i\alpha}$ contains three parts: repositioning time $t_{i\alpha}$ from region $i$ to region $\alpha$, waiting time $w_\alpha$ for getting a customer in region $\alpha$, and expected service time $\sum_{j=1}^L q_{\alpha j}t_{\alpha j}$ of transporting a customer from $\alpha$ to his destination. We assume that vehicles stay idle and do not incur driving costs while waiting. Based on that, we define the `active' driving time of a vehicle executing action $\alpha$ in state $i$ as $\tau^{dr}_{i\alpha}=t_{i\alpha}+\sum_{j=1}^L q_{\alpha j}t_{\alpha j}$, which is constant and equivalent to the time it moves on the road. But, the waiting time $w_\alpha$ depends on the total rate of vehicles that decide to take action $\alpha$ when they become empty anywhere in the system, and is defined endogenously later.

It should be noted that for taking action $\alpha$ in state $i$, AVs and CVs experience the same $\tau^{dr}_{i\alpha}$, but the expected profits the platform collects from them are different. The platform collects a portion $R\in (0,1)$ of customers' payment to CVs as commission fee\footnote{The commission that Uber takes from its drivers varies with regions and the service types. In general, the commission rate $R$ ranges from $20\%$ to $30\%$ of the fare.} and CV drivers keep the remainder, while the whole customers' payment to AVs is obtained by the platform. Besides, the driving cost of AVs is paid by the platform while the driving cost of CVs is paid by the drivers themselves. We assume a common driving cost (from fuel/electricity) rate of $c$ for each `active' AV and CV that is busy transporting customers or repositioning, and zero waiting cost for `idle' vehicles that queue in the current region for the next customer. This implies the following net profit for actions taken by AVs and CVs.

We next define the profit structure of the problem. We use the superscripts $A$ and $C$ to refer to vehicles of types AV and CV respectively, and $C2P$ to refer to revenue transferred from a CV to the platform. It must be clear that $C$ type of profit gives rise to selfish behavior for CVs, while the rest $C2P$ of the profit is collected by the platform.
Using this notation,
the expected profit an AV generates to the platform by taking action $\alpha$ in region $i$ is 
\[r_{i\alpha}^A =  p\sum_{j=1}^L q_{\alpha j}t_{\alpha j}-c\tau^{dr}_{i\alpha},\] 
where the first term is the expected revenue from the customer, and the second term is the total expected driving cost associated with action $\alpha$ in state $i$.

The expected profit a CV receives by taking action $\alpha$ in region $i$ is 
\[r_{i\alpha}^C = p(1-R)\sum_{j=1}^L q_{\alpha j}t_{\alpha j}-c\tau^{dr}_{i\alpha},\] 
where the first term is the revenue of the driver after paying the commission rate $R$ and the second term is the total expected driving cost. Then, the corresponding expected commission fee a CV generates to the platform is 
\[r_{i\alpha}^{C2P} = pR\sum_{j=1}^L q_{\alpha j}t_{\alpha j}.\]

Note our assumption that waiting cars stay idle and do not incur any driving cost allows for the profit from each action $\alpha$ to be fixed, and not depend on waiting times (which in turn depend on other drivers' decisions). Changing this assumption and assuming that vehicles move while waiting to get customers makes the cost of actions depend on waiting times. But then, the whole cost structure is simplified since there is a fixed driving cost per day that does not depend on driver decisions and driving cost becomes a sunk quantity. Our analysis methodology remains essentially the same in both cases of cost structure. 

In our model AVs are fully controlled, i.e., the platform chooses the action $\alpha$ in each region $i$, but CVs act selfishly choosing their actions to maximize average profit per unit time. Assuming that individual vehicles are too `small' to affect residual demand, i.e., the demand left for others to serve, and waiting times at the different regions. Given any amount of demand offered to CVs by the platform, the selfish choice of actions by CVs will reach an equilibrium where no CV likes to change its strategy.
Our aim in this paper is to choose the fraction of the demand to be made available to CVs and serve the remaining demand by AVs, in order to optimize the total revenue of the platform.

To make the problem more tractable, we assume that service differentiation of customers based on their destinations is not in the scope of this work, i.e., in each region $i$, the platform doesn't assign empty vehicles to arriving customers's requests depending on the customers' destinations. Only in this way the customer routing probabilities $q_{ij}$ are constant and do not depend on the policy of the platform.
This is necessary in the optimization problem for the expected rewards $r_{i\alpha}^{A}$, $r_{i\alpha}^C$, $r_{i\alpha}^{C2P}$ and transition probabilities $q_{\alpha j}$ of each vehicle taking action $\alpha$ in region $i$ to be constant and not part of the optimization problem.


\section{CV Equilibrium}
In this section, we consider the base case scenario where all the vehicles are CVs and the demand offered by the platform is $\bm{b}^C = \{b_i^C\}$, where $b_i^C\leq b_i$ for all region $i$. Given this demand, each self-interested CV driver will choose its strategy by solving its corresponding MDP that maximizes his long-run average reward assuming the parameters of the system (i.e., waiting times, cost structure, demand structure) are fixed. The key observation is that CVs interact by depleting a set of common resources, which in our case correspond to the customer demand at the various regions. These externalities are captured through the need for waiting in the vehicle queues that form in a region when supply exceeds demand. The waiting times will affect the cost structure of the MDPs and the drivers will take these into account when optimizing their strategy, which in turn will affect the size of the queues, etc. 

This collective behavior of the CVs is a special case of the mean-field game analyzed in \citealt{antonis}. Based on that, we derive for our specific ride-hailing application the basic equilibrium analysis results that characterize the behavior of the CV fleet, and then define the mixed-fleet optimization problem when we include AVs fully controlled by the platform. For a more formal definition of the game, the reader is referred to \citealt{antonis}. We refer to any equilibrium of this game as a `CV equilibrium'.

\subsection{Problem Formulation}
\label{sec:pf}
Let $x_{i\alpha}$ be the total rate at which action $\alpha$ (i.e., reposition to $\alpha$ to serve the next customer) is taken in region $i$ by the aggregate of CVs. In our optimization of driver repositioning strategies we will be optimizing over $x_{i\alpha}$s since
given any feasible state-action rates of vehicles $\bm{x} = \{x_{i\alpha}\}$, the probability of choosing action $\alpha$ in region (MDP state) $i$ is uniquely determined by $\frac{x_{i\alpha}}{\sum_{k=1}^L x_{ik}}$, i.e., any feasible set of state-action rates can be translated into state strategies (probabilities for choosing actions in states). Since CVs are homogeneous, we can also assume that all CVs use the same strategies.

We are ready now to define a CV equilibrium as a pair $(\bm{w}^C = \{w_{i}^C\},\bm{x}^C = \{x_{i\alpha}^C\})$ of CVs' waiting times and state-action rates to capture the interaction between strategic CV drivers. The platform makes available to CVs a demand vector $\bm{b}^C\leq \bm{b}$.
\begin{definition}[CV equilibrium]
\label{def:CVE}
Given a demand vector $\bm{b}^C$, a pair $(\bm{w}^C = \{w_{i}^C\},\bm{x}^C = \{x_{i\alpha}^C\})$ is an equilibrium if and only if

i) Best response: $\{x_{i\alpha}^C\}$ is the solution to the long-run average reward (profit) maximization problem of the CVs for the (fixed) waiting times $\bm{w}^C = \{w_{i}^C\}$, i.e.,
\begin{subequations}\label{equ:cvlp}
\begin{align}
\max_{\bm{x}^C} \quad & \sum_{i=1}^L \sum_{\alpha=1}^L r_{i\alpha}^C x_{i\alpha}^C \label{equ:cvprofit} \\
\text{s.t.} \quad & \sum_{j=1}^L (\sum_{k=1}^L x_{kj}^C) q_{ji} = \sum_{\alpha = 1}^L x_{i \alpha}^C, \quad \forall i, \label{equ:cvbalance}\\
& \sum_{i=1}^L \sum_{\alpha=1}^L (\tau^{dr}_{i\alpha}+w_{\alpha}^C)x_{i\alpha}^C = N, \label{equ:cvfleetsize}\\
& x_{i \alpha}^C \ge 0, \quad \forall i, \alpha.
\end{align}
\end{subequations} 

ii) Feasibility: The supply rate of CVs does not exceed the offered demand, i.e., $\sum_{j=1}^L x_{ji}^C\le b_i^C, \forall i$, and

iii) Complementary slackness: The waiting times $\bm{w}^C$ are consistent with the flow rates $\bm{x}^C$ and the offered demand $\bm{b}^C$, i.e., $w_i^C(b_i^C-\sum_{j=1}^L x_{ji}^C)=0,\forall i$.
\end{definition}
\begin{corollary}
    Part i) of the equilibrium definition characterizes the optimal behavior of a single CV assuming the system delays are fixed. 
\end{corollary}
This is because the system of equations is homogeneous in the rates, and a single CV will correspond to the solution $\bm{x}^C$ by replacing $N$ with $1$ in \eqref{equ:cvfleetsize}, i.e., act as to provide the fraction $1/N$ of the total rate. 

In the equilibrium no CV wants to change its behavior assuming the system delays as given, the total rates are feasible, and the delays must satisfy the slackness conditions. In part i) feasibility is expressed in terms of rate balance in each state, i.e., equation \eqref{equ:cvbalance}, and Little's Law (mass conservation), i.e., equation \eqref{equ:cvfleetsize}.

It turns out that any CV equilibrium can be characterized as the solution of a convex optimization problem, where the input is the demand $\bm{b}^C$ made available to CVs.

\begin{proposition}[CV equilibrium characterization]
\label{prop:NE}
 In any CV equilibrium  $(\bm{x}^C,\bm{w}^C)$ corresponding to demand vector $\bm{b}^C$, the action rates $\bm{x}^C$ solve the  convex optimization problem $\mathcal{CV}(\bm{b}^C)$ and the waiting times $\bm{w}^C$ are the Lagrange multipliers of the constraint \eqref{equ:cvdemand}, where
\begin{subequations}\label{equ:lowerlevel}
\begin{align}
\mathcal{CV}(\bm{b}^C):\quad\max_{\bm{x}^C} \quad & N\log\sum_{i=1}^L\sum_{\alpha=1}^L r_{i\alpha}^C x_{i \alpha}^C-\sum_{i=1}^L\sum_{\alpha=1}^L \tau^{dr}_{i\alpha}x_{i\alpha}^C \label{equ:objstage2}\\
\text{s.t.}  \quad & \sum_{j=1}^L x_{ji}^C \le b_i^C, \quad \forall i, \label{equ:cvdemand} \\
& \sum_{j=1}^L (\sum_{k=1}^L x_{kj}^C) q_{ji} = \sum_{\alpha = 1}^L x_{i \alpha}^C, \quad \forall i, \\
& x_{i \alpha}^C \ge 0, \quad \forall i, \alpha.\label{equ:lezero}
\end{align}
\end{subequations}    
\end{proposition}
The proof is given in Appendix \ref{app:cvx}. Note that the second term in the objective function, $m_0^C=\sum_{i=1}^L\sum_{\alpha=1}^L \tau^{dr}_{i\alpha} x_{i \alpha}^C$,
is the number of `active' CVs, i.e., \emph{not waiting in queues}. Active vehicles at any time are serving customers or repositioning. 

Note also that the KKT conditions of constraint \eqref{equ:cvdemand} guarantee that there are no CVs waiting in a region if the rate of arriving CVs is less than the rate of arriving customers, i.e., $w_i(b_i^C-\sum_{j=1}^L x_{ji}^C)=0$. This is consistent with part iii) in the definition of the CV equilibrium in Definition \ref{def:CVE}. 


\subsection{Properties of Optimal Solution of $\mathcal{CV}(\bm{b}^C)$}\label{sec:cvproperty}

The next proposition proved in Appendix \ref{app:uniqueness} summarizes some important properties of the equilibrium.

\begin{proposition}\label{pro:uniqueness}
Given demand $\bm{b}^C$, there might exist multiple CV equilibria, i.e., multiple solutions $(\bm{x}^C,\bm{w}^C)$ to the convex optimization problem $\mathcal{CV}(\bm{b}^C)$, but all such solutions 
\begin{enumerate}
\item Generate the same profit $\pi(\bm{b}^C)=\sum_{i=1}^L\sum_{\alpha=1}^L r_{i\alpha}^{C2P} x_{i \alpha}^C$ to the platform.
\item Use the same active mass $m_0^C=\sum_{i=1}^L\sum_{\alpha=1}^L \tau^{dr}_{i\alpha} x_{i \alpha}^C$ of CVs.
\item $\pi(\bm{b}^C)$ obtained by solving $\mathcal{CV}(\bm{b}^C)$ is non-decreasing with CV fleet size $N$.
\item For any CV equilibrium $(\bm{x}^C,\bm{w}^C)$ obtained by solving $\mathcal{CV}(\bm{b}^C)$, the active mass $m_0^C$ of CVs acts the same as if centrally controlled by the platform.
\end{enumerate}
\end{proposition}
The proof is given in Appendix \ref{app:uniqueness}. To make the last statement more precise, the vehicle repositioning activity $x_{i \alpha}^C$ observed in the equilibrium is the same as in the platform profit maximization problem when the platform has \emph{full control} over a mass of $m_0^C$ CVs (i.e., same as the mass of active CVs in the equilibrium). This corresponds to
\begin{subequations} \label{equ:controlcv}
\begin{align}
\max_{\bm{x}^C} \quad & \sum_{i=1}^L\sum_{\alpha=1}^L r_{i\alpha}^C x_{i \alpha}^C \label{equ:objstagenew}\\
\text{s.t.}  \quad & \sum_{j=1}^L x_{ji}^C \le b_i^C, \quad \forall i, \label{equ:demand2}\\
& \sum_{j=1}^L (\sum_{k=1}^L x_{kj}^C) q_{ji} = \sum_{\alpha = 1}^L x_{i \alpha}^C, \quad \forall i,\\
&\sum_{i=1}^L\sum_{\alpha=1}^L \tau^{dr}_{i\alpha}x_{i\alpha}^C\leq m_0^C, \label{equ:activem}\\
& x_{i \alpha}^C \ge 0, \quad \forall i, \alpha.
\end{align}
\end{subequations} 

Although the platform profit is monotone with respect the mass of CVs, it is not monotone in the demand of customers (which we expect to be the case if vehicles are fully controlled).

\begin{observation}\label{pro:nomonotone}
There are examples of systems where it is possible for the platform to increase profit by hiding demand from CVs, i.e., $\pi(\bm{b}^C)$ can be decreasing in $b_i^C$ for some $i$. 
\end{observation}

As discussed in \citealt{purecv}, the intuition behind this observation \footnote{This will also be discussed in the analysis of the general two-region network that is given in Appendix \ref{app:tworegion}.} is that the platform can incentivize CVs that queue in a given region to reposition by reducing the demand in that region and increasing the corresponding waiting time. Due to the increased waiting, a fraction of the CVs will find it more profitable to reposition and serve other regions, resulting in more customers being served overall compared to the previous situation. 

We prove that this cannot be the case if there is no CV queuing in the equilibrium. 
\begin{proposition}\label{pro:nohiding}
If $\bm{w}^C=\bm{0}$ in the equilibrium $\mathcal{CV}(\bm{b}^C)$, it is never profitable for the platform to hide demand from CVs. 
\end{proposition}

The proof is given in Appendix \ref{app:nohiding}.

\subsection{The Equilibrium As A Choice of Policy Cycles}
\label{sec:pcycles}

We like to offer an alternative way to understand the CV equilibrium, that can help the intuition of the reader to understand some of the examples analyzed later. We perform the discussion at an informal level because it does not play a central role in the results of this paper. 

Given any region graph, define as a \emph{policy cycle} a deterministic choice of repositioning actions in each region. This corresponds to a Markov chain, since after repositioning to a given region, the next transition is defined by the routing choice of the customer which is fixed.
We call this a `cycle' since it renews itself each time a reference region is visited. There are exponentially many cycles, but let's not worry about complexity. Given a set of waiting times $\bm{w}$ in the customer queues of the different regions, for any given cycle $s$ we can compute its expected execution time $T_s$, (i.e. the expected time between subsequent visits to a reference region). Each cycle $s$ has a unique value of average reward $r_s$ that depends on the average traveling time it will visit the various regions and the average reward per region. Hence, given $\bm{w}$, a cycle $s$ generates an average rate of reward $\bar{r}_s=\frac{r_s}{T_s(\bm{w})}$ that depends on the queuing delays in the regions.

CV driver choices correspond to choosing cycles. For any allocation of the driver mass to cycles, there is a corresponding contention for resources (customers) and a corresponding set of $\bm{w}$. This defines for each cycle $s$ its average reward rate $\bar{r}_s$, and drivers will try to move to cycles with the highest such reward rate. In doing so, popular cycles will incur more resource congestion, their waiting times will increase, thus reducing their reward rate. The CV equilibrium corresponds to the case where all cycles that have positive driver mass generate the same reward rate.

Thinking of drivers choosing cycles according to their reward rate helps our intuition to understand how equilibria are formed. Assume a mass of $N$ drivers. Start adding drivers to the system starting from zero mass. The first driver will choose a cycle with the highest value of $\bar{r}_s$. As more drivers join this cycle, queues will start forming in certain regions and $\bar{r}_s$ will start to decrease. Eventually, we may get a situation where the cycle $s$ still remains with the highest reward rate among all the other cycles even after adding all $N$ drivers, or it starts to become equal to a cycle with the second-highest reward rate as queues reach some specific length. Note that reward rates of other cycles don't stay constant and may also drop due to the increase of the waiting times at the different regions from the congestion caused by the first cycle. In the later case, new drivers will be added 
to these two highest reward cycles and existing drivers may have to move among cycles in a way to keep these cycles at the same reward rate level. As more drivers are added, more cycles might be added to the optimum set by matching the reward rate of the cycles already in the optimal set. Eventually, all cycles with positive activity will have the (same) highest reward rate among all the possible cycles (all other cycles will offer a reward rate that is lower or equal). We make these ideas more concrete through the following simple example. 

\begin{figure}
    \centering 
    \includegraphics[width=6cm]{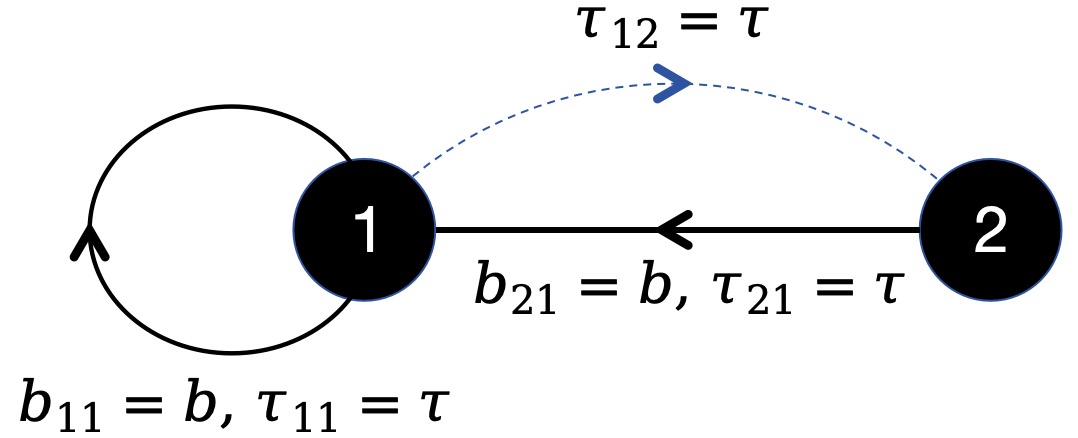}
    \caption{
    A two-region ride-hailing network where all customers have the same destination region 1. The demand rate is $b_{11}=b_{21}=b$ (black arrows), $b_{12}=b_{22}=0$, and the trip time of any origin-destination pair is $\tau$, implying the same payment per trip. Vehicles need repositioning from region 1 to region 2 (blue dashed arrow) to serve demand $b_{21}$ and spend an additional time $\tau$.}
    \label{fig:l2new}
\end{figure}

\noindent\textbf{Example 1.} Our ride-hailing network has $L=2$ regions as shown in Fig.~\ref{fig:l2new}, where all the customers originating from regions 1 and 2 want to travel to region 1, e.g., region 1 is the CBD, region 2 is the residential area, and we consider the morning traffic when all customer trips are destined to the CBD. Assume the demand rates $b_{11}=b_{21}=b$ and $b_{12}=b_{22}=0$, equal traveling times within and across regions $\tau_{11}=\tau_{21}=\tau_{12}=\tau$ and driving cost rate $c=0$ (for simplicity). Since trips have the same duration, each customer trip generates the same payment (say $p\tau=1$\$ per trip). Vehicles become free only in region 1 and need repositioning to region 2 (blue dashed arrow) to serve demand $b_{21}$. 

There are four possible policy cycles: for each region $i$, choose to serve the next customer in region $j\in \{1,2\}$. In our case, since there is no demand traveling to region 2, these reduce to two different policy cycles: always serve new customers from region 1 (cycle 1), or always serve new customers from region 2 (cycle 2). Cycle 1 corresponds to always cycling around region 1 and the reward rate (in $\$$ per unit time) is $\frac{1}{\tau+w_1}$. Cycle 2 corresponds to always cycling between region 1 and region 2, and the reward rate is $\frac{1}{2\tau+w_2}$. If no queues are formed in any of the regions, cycle 1 generates twice the reward rate of cycle 2 since the time to complete cycle 1 is $\tau$ instead of $2\tau$ for cycle 2, and both produce the same reward of $1\$$ upon completion. 

Before we proceed to analyze this example, let's review Little's law. Consider an activity that takes time $\tau$ to perform, can be performed at some maximum rate $b$, and there is a waiting room for agents to wait before they can have access to the activity. Assume that the total mass $m$ of agents that are involved at any time in this activity (either performing the activity or waiting) is fixed, $w$ is the time spent in the waiting room, and $x$ is the rate of performing the activity. Little's law implies the unique characterization of $x,w$ as follows,
\begin{equation}
    m=x(\tau+w), x\leq b, w(b-x)=0\,.
    \label{eq:ll}
\end{equation}
 As $m$ increases from zero, $x$ increases proportionally to satisfy $m=x(\tau+0)$ with $w=0$. When $x$ reaches the maximum activity rate $b$, it cannot increase any longer and then $w$ starts increasing according to $m=b(\tau+w)$.
Hence, there is no waiting $w=0$ if $x\leq b$ or equivalently if $m\leq b \tau$. Only when the mass of agents is larger than $b\tau$ there will be queuing, and $w=\frac{m}{b}-\tau$.

Let's see now how CV equilibria will be formed as we increase the mass of drivers.
When the system is empty of drivers there is no queuing, i.e., $w_1=w_2=0$, and the reward rates of the cycles are $r_1=\frac{1}{\tau}=2r_2$, since $r_2=\frac{1}{2\tau}$.  As we add more drivers, these choose the cycles with the largest rate of reward. We can distinguish the following types of equilibria depending on the total mass $m$ of drivers.

\begin{itemize}
    \item All drivers join cycle 1, with $w_1=w_2=0$. This is the case when $m\leq \frac{b}{\tau}$. There is no queuing in region 1 and $r_1=2r_2$.
    \item All drivers join cycle 1, with $w_1>0,w_2=0$. In this case by Little's law $m=b(\tau+w_1)$, and $w_1$ must be small enough in order for cycle 1 not to become less profitable than cycle 2, i.e., $\frac{1}{\tau+w_1}\geq \frac{1}{2\tau}$. This holds when $w_1\leq \tau$, implying $m\leq b2\tau$. When $m = b2\tau$, both cycles generate the same reward rate.
    \item Drivers join both cycles, with $w_2=0$. If we increase $m$ further from $m = b2\tau$, all the new drivers will join cycle 2. This is because i) there will be no queuing initially in region 2 and $r_2$ will remain constant and equal to $\frac{1}{2\tau}$ matching the reward of cycle 1, while ii) if more drivers join cycle 1, this will strictly increase $w_1$ and hence, decrease $r_1$.  Using similar arguments, the additional mass of $b2\tau$ can be added to cycle 2 before $w_2>0$. Hence, queuing in region 2 will not start while the total mass is $m \leq 4b\tau$.
    \item When $m > 4b\tau$, driver mass in excess of $4b\tau$ will join both cycles, increasing $w_1,w_2$ in a way that $\frac{1}{\tau+w_1} = \frac{1}{2\tau+w_2}$. 
\end{itemize}

We can make the following interesting observations. When the mass of drivers is small,  $m\leq \frac{b}{\tau}$, the system is efficient and the equilibrium is socially optimal and also maximizes the platform profit since drivers serve the maximum possible demand, the same as if they were under the direct control of the platform. The same holds when the mass is large $m > 4b\tau$, because all demand is also served. But for intermediate values of $m$, in the equilibrium the drivers serve fewer customers than they would otherwise serve if they were under direct control of the platform. For example, when $m=2b\tau$, a mass $b\tau$ is `useful' by being actively involved in fully serving the demand $b$, and a mass $b\tau$ is involved in waiting, generating no revenue for the platform. The platform would have preferred these drivers instead of waiting to serve cycle 2. They would serve an amount $b/2$ of demand from region 2 and thus generate additional reward to the platform.

The next discussion is related to how to deploy a single fully controlled vehicle (an AV), in the simple case that it generates almost the same profit as a selfish vehicle (a CV), i.e., $R=1$. Suppose the platform wants to deploy an AV additional to the existing CV fleet. Where should it be deployed? In cycle 1 or 2? The answer depends on the current CV equilibrium. If we are in type 1 equilibrium, deploying the AV to the most rewarding cycle is best, since there is residual demand to be served. But if we are in type 2 equilibrium, deploying the AV in cycle 1 might not be the optimal choice since it will add no additional revenue: it will only increase the number of queuing CVs by one, serving the same demand as before. But if the AV is deployed in cycle 2 instead, this will serve additional demand and generate additional revenue. If the equilibrium is of type 3, deploying the one AV in cycle 1 will induce two CV drivers to move to cycle 2, serving more demand than if the AV was initially dispatched to cycle 2. In the case of type 4 equilibrium, adding an AV has no effect in the revenue of the platform \footnote{Actually, it will have an effect when $R<1$ since it will displace a CV. If $R=1-\epsilon$, this will be an additional profit $\epsilon$ generated by replacing a CV with the AV.}. This discussion shows the complexity of controlling a mixed fleet, which is the objective of the next section.

\section{Mixed-fleet Problem Formulation}\label{sec:twostage}
\label{sec:mixedfleet1}


In this section, we study the mixed-fleet platform profit maximization problem where there are both AVs controlled by the platform and CVs operated by selfish drivers. This is a bi-level optimization problem where at the first level the platform chooses how to operate its AVs and the demand allocated to CVs, and at the second level the CVs optimize their strategies by taking into account the resulting conditions from their actions and the actions of the platform. In the corresponding (Stackelberg type of) equilibrium, the platform does not want to change its actions, assuming that  CVs will always react by optimizing their own profit.

Our mixed-fleet equilibrium is characterized by a $4$-tuple $(\bm{x}^A, \bm{w}^A,\bm{x}^C,\bm{w}^C)$, corresponding to sets of state-action rates and waiting times for AVs and CVs. We don't know how this equilibrium is achieved, but we require that it satisfies three properties: i) feasibility (i.e., capacity constraints, flow balance constraints and mass conservation, including both AVs and CVs), ii) the CVs are in equilibrium, i.e., self-interested CV drivers maximize their long-run average reward assuming the system conditions as fixed, and iii) a `platform equilibrium', i.e., the platform maximizes its profit under all situations that satisfy i) and ii). 
 
\begin{definition}[Mixed-fleet equilibrium] \label{d:mixed equi}
In the mixed-fleet ride-hailing system with demand $\bm{b}$, $(\bm{x}^A, \bm{w}^A,\bm{x}^C,\bm{w}^C)$ is an equilibrium if and only if

i) Feasibility: capacity constraints, flow balance constraints and mass constraints are satisfied:
\[\sum_{j=1}^L (x_{ji}^A+x_{ji}^C) \le b_i , \quad \forall i,\]
\[\sum_{j=1}^L (\sum_{k=1}^L x_{kj}^V) q_{ji} = \sum_{\alpha = 1}^L x_{i \alpha}^V, \quad \forall i,V\in\{A,C\} \]
\[\sum_{i=1}^L\sum_{\alpha=1}^L (\tau^{dr}_{i\alpha}+w_{\alpha}^A) x_{i\alpha}^A \le M,\quad \sum_{i=1}^L\sum_{\alpha=1}^L (\tau^{dr}_{i\alpha}+w_{\alpha}^C) x_{i\alpha}^C = N.\]

ii) CV equilibrium: No CV wants to deviate, i.e., $(\bm{x}^C,\bm{w}^C)$ satisfy the conditions of part i) in Definition \ref{def:CVE}.  

iii) Platform equilibrium: the total platform profit $\sum_{i=1}^L\sum_{\alpha=1}^L r_{i\alpha}^A x_{i\alpha}^A +\sum_{i=1}^L\sum_{\alpha=1}^L r_{i\alpha}^{C2P} x_{i\alpha}^C$ generated by AVs and CVs is maximized over all action rates and waiting times of AVs and CVs that satisfy conditions i) and ii).  
\end{definition} 

Observe that in i) the constraints are symmetric for AVs and CVs, except for the mass constraints. This is because the platform can choose to use only part of the AVs, leaving the rest in storage. 
Note also that we don't impose any complementary slackness conditions with waiting times and vehicle service rates $x_i^C$, $x^A_i$ of CVs and AVs in each region $i$. Since we allow for arbitrary customer service policies, possibly not work conserving, we could have $w^C,w^A>0$ even if $x_i^C+x^A_i<b_i$.

Let $\bm{b}_e^C=\{\sum_{j=1}^L x_{ji}^C\}$ be the demand served by CVs in a mixed-fleet equilibrium, and $\mathcal{CV}(\bm{b})$ defined in Proposition \ref{prop:NE}.
\begin{proposition}
In any mixed-fleet equilibrium $(\bm{x}^A, \bm{w}^A,\bm{x}^C,\bm{w}^C)$, the CV rates-waiting times $(\bm{x}^C,\bm{w}^C)$ must solve $\mathcal{CV}(\bm{b}_e^C)$.
\end{proposition}
The proof is straightforward. It follows the same steps as the proof for the pure CV case. 

The profit of any equilibrium in which $w^A>0$, can be obtained in an equilibrium where no AV is waiting in queues and CVs see the same queuing conditions as before.  The next lemma simplifies the form of equilibria that we try to analyze.
\begin{lemma}  \label{l:active}
For any mixed-fleet equilibrium $(\bm{x}^A, \bm{w}^A,\bm{x}^C,\bm{w}^C)$, there exists an equivalent equilibrium $(\bm{x}^A, \bm{0},\bm{x}^C,\bm{w}^C)$ that generates the same platform profit. 
\end{lemma}
The proof is simple. Serve CVs as before. Transfer the mass of waiting (i.e., idling) AVs in storage and keep utilizing only the `active' AVs. This will not change the service rate by the AVs and the amount of served demand. Hence, we can obtain the same revenue by having AVs either fully active or unutilized.



We next provide an `operational' characterization of a mixed-fleet equilibrium.
\begin{corollary}
    Any mixed-fleet equilibrium $(\bm{x}^A, 0,\bm{x}^C,\bm{w}^C)$ can be operationally induced by the following simple platform mechanism: the platform reveals to CVs demand equal to $\bm{b}_e^C$ (the demand that was served in the equilibrium by CVs). Then, serves the remaining demand $\bm{b}_e^A=\bm{b}-\bm{b}_e^C$ using AVs as dictated by $\bm{x}^A$. 
\end{corollary}

This suggests the following optimization problem to attain the profit of any mixed-fleet equilibrium. The decision variables are the demand allocated to CVs and the AV service rates for the remaining demand. 

\begin{definition}[Bi-level optimization]
    \label{d:BLO}
\begin{subequations}\label{equ:bilevel}
\begin{align}
\mathcal{OPT}: \quad \max_{\bm{b}^C,\bm{x}^A} \quad & \sum_{i=1}^L \sum_{\alpha=1}^L r_{i\alpha}^A x_{i\alpha}^A + \pi(\bm{b}^C) \label{equ:obj2}\\
\text{s.t.}  \quad & \sum_{j=1}^L x_{ji}^A+b_i^C \le b_i , \quad \forall i, 
\label{equ:demand}\\
& \sum_{j=1}^L(\sum_{k=1}^L x_{kj}^A) q_{ji} = \sum_{\alpha}x_{i\alpha}^A \quad \forall i, \label{equ:avbalance}\\
&\sum_{i=1}^L\sum_{\alpha=1}^L \tau^{dr}_{i\alpha}x_{i\alpha}^A \le M, \label{equ:mav} \\
& b_i^C, x_{i \alpha}^A \ge 0, \quad \forall i, \alpha. \label{equ:range}
\end{align}
\end{subequations} 

\end{definition}
The objective function \eqref{equ:obj2} is the platform profit earned by AVs plus the commission fees collected from CVs operating in equilibrium. Note that the CVs are fed an amount of demand $\bm{b}^C$ and their equilibrium is reached by forming their own waiting queues to serve that demand. The constraints \eqref{equ:demand}-\eqref{equ:mav} here are similar to those in \eqref{equ:lowerlevel}, including the capacity constraint and flow balance constraint in each region. In the mass constraint \eqref{equ:mav} we express the fact that no waiting is necessary for AVs to optimally serve the given demand, allowing AVs to remain unutilized. 

The reason the above bi-level optimization problem can maximize the profit of the platform over all possible platform strategies is because it can choose to hide demand from CVs even if this is not served by AVs. This is expressed in \eqref{equ:demand} that does not have to be tight. We may restrict $b_i^C$ to be strictly less than the demand unserved by AVs. \emph{This is not the case if we change $\mathcal{OPT}$ to first decide in the demand $\bm{b}^A$ to be allocated to AVs and then release all the remaining demand to CVs}. 

In the next theorem, we formally establish the (obvious by now) connection between the bi-level optimization problem and mixed-fleet equilibria. 
\begin{theorem}\label{thm:bilevel}
The following hold:
\begin{enumerate}
    \item If $(\bm{x}^A,\bm{0},\bm{x}^C,\bm{w}^C)$ is a (Stackelberg) equilibrium of the mixed-fleet system, then $(\bm{x}^A,\bm{b}_e^C)$ solves $\mathcal{OPT}$ and $(\bm{x}^C$, $\bm{w}^C)$ are the optimal CV rates and the Lagrange multipliers of the capacity constraints in $\mathcal{CV}(\bm{b}_e^C)$.
    \item If $(\bm{x}^A,\bm{b}^C)$ solves $\mathcal{OPT}$ and $(\bm{x}^C,\bm{w}^C)$ are the optimal CV rates and the Lagrange multipliers of the capacity constraints in $\mathcal{CV}(\bm{b}^C)$, then $(\bm{x}^A,\bm{0},\bm{x}^C,\bm{w}^C)$ is a (Stackelberg) equilibrium of the mixed-fleet system.
\end{enumerate}
\end{theorem}


The platform profit obtained in the optimal solution of $\mathcal{OPT}$ is non-decreasing with the AV fleet size $M$ since at least the same profit can be achieved by keeping the additional AVs unutilized (remember that so far $M$ is an exogenous parameter; the endogenous choice of $M$ is discussed in later sections). Moreover, based on the monotonicity of $\pi(\bm{b}^C)$ in CV fleet size $N$, it is straightforward for us to prove the total platform profit is also non-decreasing with $N$ in the optimal solution.  

\begin{lemma}\label{lem:monotone}
In the optimal solution of the bi-level optimization problem, the platform profit is non-decreasing with AV fleet size $M$ and CV fleet size $N$.
\end{lemma}

The challenge in solving $\mathcal{OPT}$ is that we do not have $\pi(\bm{b}^C)$ in closed form. Moreover, based on the Observation \ref{pro:nomonotone}, $\pi(\bm{b}^C)$ is not monotone in $\bm{b}^C$ and thus $\mathcal{OPT}$ is a non-convex problem and can have many local maxima. One possible way to find the optimal $\bm{b}^C$ is to use an exhaustive search, but the computation is prohibitive, especially when the number of regions $L$ is large.

In the following Section \ref{sec:ana}, we focus on discussing the properties of $\mathcal{OPT}$. Then, in Section \ref{sec:heu} we propose three efficient algorithms for solving $\mathcal{OPT}$ that can achieve near-optimal performance with low complexity.

\section{Properties of $\mathcal{OPT}$}\label{sec:ana}

In this section, we discuss some interesting properties of the optimal solution $\bm{b}^{C\star}$, $\bm{x}^{A\star}$ and $\bm{x}^{C\star}$ of the bi-level optimization problem $\mathcal{OPT}$ for the mixed-fleet system.

\subsection{Fully Active AVs}
As we argued in Lemma \ref{l:active}, there is no reason for a platform to maintain idle AVs in queues, and AVs will be either active or unutilized.
Let $m_{0}^{A\star}=\sum_{i=1}^L \sum_{\alpha=1}^L \tau^{dr}_{i\alpha}x_{i\alpha}^{A*}$ be the mass of active AVs used in the optimal solution $\bm{x}^{A\star}$ of the bi-level optimization problem. Clearly, $m_{0}^{A\star}\leq M$.


\begin{definition}
A `fully active' AV strategy is a strategy under which all AVs are active.
\end{definition}




We argue in the next proposition that it is inconsistent for the platform to offer demand to CVs if there are unutilized AVs, and that if there are enough AVs to fully serve the demand, then there is no reason to deploy CVs. The intuition is that if there is unserved demand by AVs that is served by CVs while there are unutilized AVs, the platform could use such AVs to simulate the behavior of CVs and increase its profit. This is due to Lemma \ref{lem:monotone} and the fact that AVs generate more profit per trip to the platform compared to CVs. 

\begin{proposition}\label{pro:allav}
In the optimal solution of the bi-level optimization problem, if there are CVs serving customer demand, i.e., $\bm{x}^{C\star}\neq \bm{0}$, then the following must be true
\begin{enumerate}
    \item The optimal AV strategy $\bm{x}^{A\star}$ is `fully active', i.e., $m_{0}^{A\star} = M$.
    \item There exists no AV strategy that can serve all customer demand. 
\end{enumerate}
\end{proposition}

The proof of Proposition \ref{pro:allav} is given in Appendix \ref{app:allav}.

\subsection{AV-first Policy}
From the above analysis one may wonder if the optimal AV strategy $\bm{x}^{A\star}$ can be obtained by independently optimizing the AV usage without considering CVs. We formally define this simple heuristic as the `AV-first' policy.

\begin{definition}
    \textit{AV-first Policy}: Deploy the available AVs to maximize platform profit without considering the existence of CVs. If there is any residual demand, then reveal it to CVs. 
\end{definition}
The AV-first strategy $\hat{\bm{x}}^A$ is the solution of problem $\mathcal{OPT}$ by setting $\bm{b}^C = \bm{0}$ (all demand available to AVs), and $\hat{\bm{x}}^C$ is the solution of problem $\mathcal{CV}(\bm{b}^C)$ by substituting $b_i^C=b_i - \sum_{j}\hat{x}_{ji}^A$ (all the unserved demand from AVs available to CVs).

Since for each $1\$$ generated by an AV, a CV generates $R\$$ and $R<1$, one might expect that the AV-first policy maximizes the platform profit because it deploys AVs as efficiently as possible to serve regions that generate higher rates of revenue by requiring small repositioning costs. Since CVs are less profitable to the platform, it should be sensible to allocate to them demand from less profitable regions.  

Although this should be the case when $R$ is very small, for larger values of $R$ we will see through a representative example that this might not be the case. In this example, the platform will prefer to offer to CVs the demand that generates the highest rate of revenue and use the AVs to serve demand that CVs did not find profitable to serve in the AV-first allocation. This increase of total served demand, although served partly by CVs that keep part of the reward for themselves, can lead to an increase of the total profit of the platform. The crucial issue is to improve the overall efficiency of the system by reducing the useless waiting of CVs in queues, and have them serve more customers instead. To achieve that, the platform has to do a smarter scheduling of its AVs since it cannot directly control the decisions of CVs. We make this intuition more concrete by returning to Example 1 (in Figure \ref{fig:l2new}).

\textit{When is AV-first optimal?}
Consider the two extreme cases where i) there is shortage of AVs and CVs, and the total mass of AVs and CVs that cannot fully serve region 1, i.e., $M+N\leq b\tau$, and ii) there is abundance of AVs so that they can serve at least region 1 fully, i.e., $M\geq b\tau$, but together they cannot serve the total demand, i.e., $M+N\leq b\tau+b2\tau$. In both cases, AV fist is optimal since it induces CVs to do the same actions as if they were fully controlled by the platform. In case 1 it will assign the AVs in cycle 1, i.e., in region 1 to serve an amount $\frac{M}{\tau}$ of demand, and in the resulting CV equilibrium the CVs will also choose cycle 1 and serve $\frac{M}{\tau}$ leaving some demand unserved, i.e., without any queuing, which is optimal for the platform. 
In case ii), the AV-first will fully serve cycle 1 by utilizing all the necessary AVs to serve region 1's demand, and allocate the rest in cycle 2 to always serve region 2. The CVs will form an equilibrium where they will choose cycle 2 (since cycle 1 has no more available demand for them) and serve demand from region 2 without forming any queues. This is again optimal for the platform.


\textit{When is AV-first not optimal?}
In both of the above cases where AV-first was optimal, under AV-first there was no CV idling, i.e., all CVs were fully utilized in serving demand. AV-first might not be optimal when CVs are idling in queues in the equilibrium, since there might be \textit{incentive incompatibility}: the platform would like idle CVs to relocate and serve more customers, while selfish drivers prefer to wait instead of relocating. When AV-first results in this incentive incompatibility situation, the platform should schedule AVs differently.

Following this reasoning, we construct a case where under AV-first a large fraction of CVs ends-up waiting in region 1 instead of relocating to region 2 where there is unserved demand. The platform can do better by allocating region 1 to CVs and using the AVs to serve region 2. This is because the CVs will always prefer to serve region 1 and will form an equilibrium where region 1 is fully served without waiting (i.e., optimally) by the CVs, and AVs can be used to serve the demand in region 2. It is straightforward to construct the right system parameters as follows.

Use a mass $M=0.5b\tau$ of AVs and $N=b\tau$ of CVs. AVs alone can serve half the demand in region 1, 
whereas CVs alone can serve exactly all the demand in region 1 without creating any waiting. If we use AV-first, AVs will serve in region 1 $0.5b$ units of demand and leave $0.5b$ to the CVs. All CVs will concentrate only to region 1 (i.e., cycle 1) to serve the remaining $0.5b$ units of demand, and generate a waiting $w=\tau$ in CV queues (since by Little's law $N=0.5b(w+\tau)$ and $N=b\tau$). This is the only possible CV equilibrium since their revenue rate is one customer payment every $2\tau$, same as if relocating to serve region 2 (cycle 2). Thus, the rate of revenue of the platform under AV-first is $0.5b+0.5bR$\$ (since from every customer payment served by a CV the platform keeps a fraction $R$).

Can we improve that? Suppose the platform offers all the demand in region 1 to CVs. In this case, all the demand in region 1 will be served, generating a reward rate $bR$. All the AVs will be sent to serve demand in region 2, in which case they will serve a rate $0.5b\tau/2\tau=0.25b$ of additional customers compared to the AV-first, generating a $0.25b$\$ rate of revenue. In total, the revenue rate is $0.25b+bR$ compared to $0.5b+0.5bR$ under AV-first. Simple calculations suggest that AV-first is not preferred if and only if $R>1/2$. If commission is high enough, the platform prefers a policy to serve more total demand, where CVs end up serving more demand than before. 

Further detailed discussion can be found in Appendix \ref{app:example1}.

In the next proposition, we provide some sufficient optimality conditions of the AV-first policy. This result applies to general networks with any arbitrary number of regions.


\begin{proposition}\label{pro:allav3}
The AV-first policy is optimal if: i) under the AV-first policy, the residual demand $\{b_i^C\} = \{b_i - \sum_{j}\hat{x}_{ji}^A\}$ is fully served by CVs in the CV equilibrium, or ii) there exists an AV policy that can serve all the demand\footnote{This can be different from the AV-first policy since the AV-first policy might not serve all the demand even when there are enough AVs to do so. This happens when, for example, some remote regions are not profitable to serve due to a large repositioning cost.}.
\end{proposition}
The proof is given in Appendix \ref{app:allav3}.

\subsection{Performance Guarantee of AV-first}
\label{sec:pg}
Due to its simplicity, AV-first makes lots of practical sense. The key question is how much worse it can be compared to the optimal policy. We have performed extensive experiments for many examples of networks, and the results suggest that the maximum loss of AV-first can be at most $20\%$ of the optimal in the extreme case of commission rate $R=1$.
$R=1$ makes CVs as lucrative as AVs, and the effect of CV repositioning to the platform profit is most important. 
We have formally proved the case of arbitrary two-region networks with cost rate $c=0$ and provided a comprehensive numerical analysis of the performance of the AV-first for three-region networks and any value of $c$. Note that $c=0$ is consistent with the case where vehicles are always moving, even when waiting to get assigned a customer, and the daily driving cost is constant becoming a sunk quantity.


\begin{proposition}\label{pro:av-first loss}
In any two-region network, the maximum performance loss of the AV-first policy is $20\%$ as compared to the optimum when driving cost rate $c=0$.
\end{proposition}

Let's first check this in the special case of Example 1 when AV-first is not optimal, specifically when $M=0.5 b\tau$ and $N=b\tau$. The reward rate is $0.5b+0.5bR$ under the AV-first and $0.25b+bR$ under the optimal policy. By comparing these two rates, it is easy to see the maximum performance loss $20\%$ of the AV-first is achieved when $R\to 1$.

Let's consider now the more general two-region networks with arbitrary arrival rates where we assume $b_{12}\leq b_{21}$ without loss of generality. Following the idea of `policy cycle', for the general two-region networks, there are two (out of the four possible) important policy cycles that we need to consider. 
Similar as in Example 1, more customers can be served by replacing in cycle 1 AVs with CVs that would otherwise be waiting and sending the AVs to serve cycle 2. As the commission rate $R\to 1$, we can show such a replacement can perform at most $20\%$ better than the AV-first. We refer readers interested in the detailed analysis of policy cycles of the general two-region networks and the rigorous proof of Proposition \ref{pro:av-first loss} to Appendix \ref{app:av-first loss}.

For networks with the number of regions $L>2$, the decomposition of demand as served by policy cycles remains feasible. However, the number of possible cycles scales up to $L^L$ (exponentially) since at each state/region an empty vehicle can choose any region for the next customer. Our numerical analysis suggests the following conjecture.

\begin{conjecture}\label{conj:gap}
For any network with an arbitrary number of regions, under the assumption of $c=0$, the maximum performance degradation of the AV-first policy is $20\%$ compared to the optimal solution.
\end{conjecture}

To substantiate the conjecture, we conducted extensive experiments for three-region networks where the optimal solution can be derived by exhaustive search. We generate a collection of $100$ instances of three-region networks. For each instance, the demand of each origin-destination pair is uniformly sampled from set $\{0,1,2\}$, the fleet sizes $M$ and $N$ are uniformly sampled from set $\{0,1,...,10\}$, and the commission rate $R$ takes a random value in the range $(0,1)$. We set the traveling time $\tau_{ij}=1$ for all origin-destination pairs $(i,j)$, customer payment per unit travel time $p=1$, and vehicle driving cost rate $c=0$. Through our simulations, we observe that the performance loss of the AV-first policy as compared to the optimum never exceeds $20\%$, and this is a tight bound.

Additionally, our simulation results demonstrated later in Section \ref{sec:numerical} suggest that with any value of driving cost rate $c$, the conclusion of the performance loss of the AV-first as compared to the optimum never exceeds $20\%$ still holds.

\section{Algorithms for Solving $\mathcal{OPT}$}\label{sec:heu}

Solving the bi-level optimization problem $\mathcal{OPT}$ 
for general $L$-region networks is a challenging task since the problem is not convex and the number of variables $\{x_{i\alpha}\}$ grows exponentially in $L$, see the following section where we discuss more on the problem structure.
In the next sections, we propose three optimization algorithms to solve $\mathcal{OPT}$ for networks with arbitrary number of regions: i) a gradient-descent algorithm that searches for a local maximum in the direction of the gradient; ii) a bundle method algorithm designed for nonsmooth optimization problems that leverages subgradients from preceding iterations; and iii) a genetic algorithm that adaptively search the solution inspired by the evolutionary process of natural selection. These algorithms, although not guaranteed to find the global optimum, present viable strategies for the platform operator to solve $\mathcal{OPT}$. We also obtain numerically interesting trends of the (approximate) optimum solution when we vary key system parameters. 

\subsection{Optimization Structure of $\mathcal{OPT}$}

\begin{figure}
    \centering 
    \includegraphics[width=15cm]{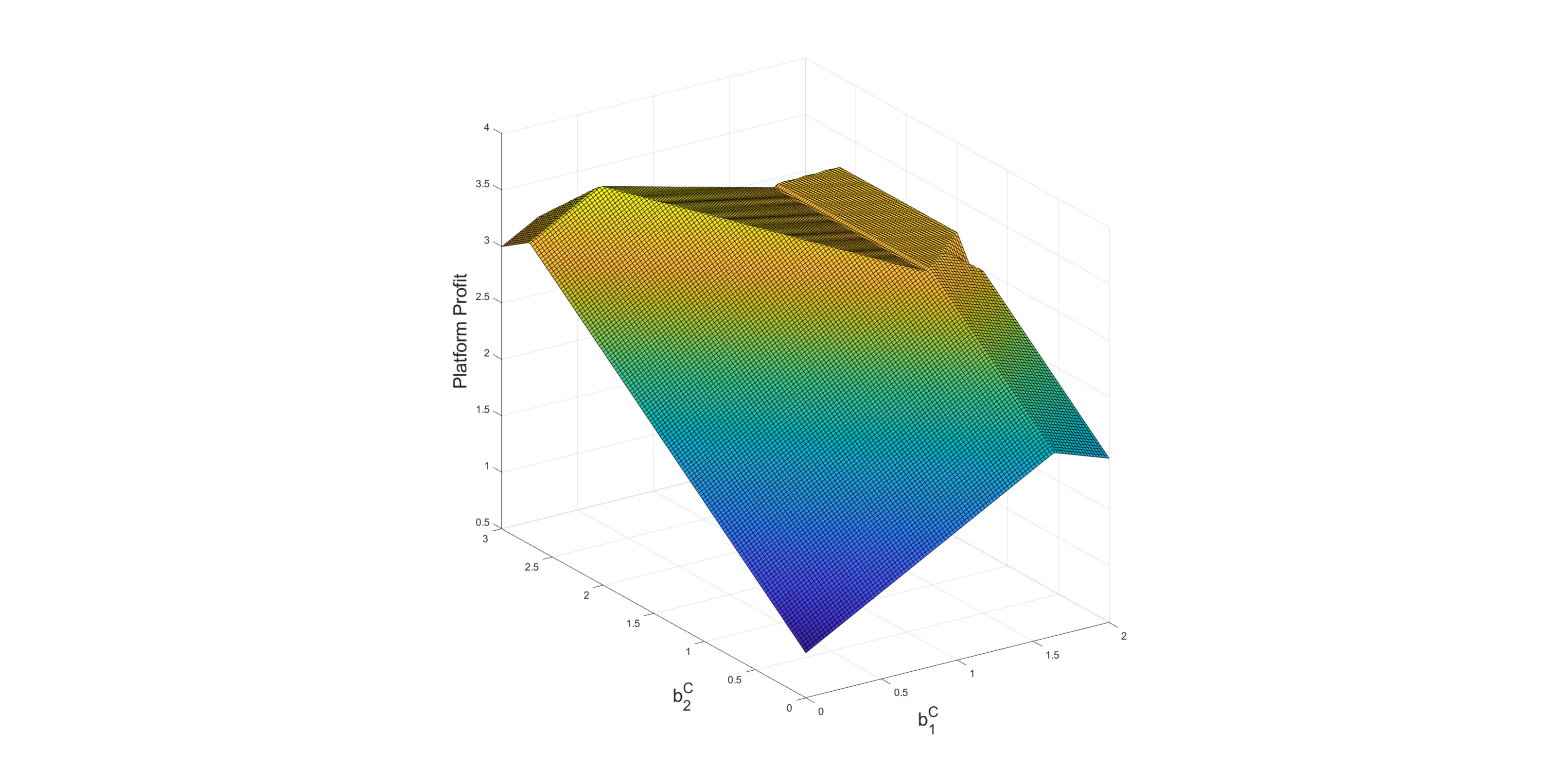}
    \caption{The platform profit value against the demand revealed to CVs for the two-region network in Fig \ref{fig:figureL2} with customer payment per unit travel time $p=1$, driving cost rate $c=0.1$ and commission rate $R=0.5$.}
    \label{fig:exhaustive}
\end{figure}

Before introducing algorithms to solve $\mathcal{OPT}$, let's first take a close look at its structure. We consider a two-region network with four customer routes, demand rates  $\{b_{11}=1, b_{12}=1, b_{21}=2, b_{22}=1\}$ and travel times $\{\tau_{11}=1, \tau_{12}=2, \tau_{21}=2, \tau_{22}=1\}$. In Fig. \ref{fig:exhaustive} we do a 3D surface plot to show how the platform profit changes as the revealed demand $b_1^C$ and $b_2^C$ to CVs in regions 1 and 2 given serving the rest of the demand $\bm{b}-\bm{b^C}$ optimally by AVs. We assume the customer payment per unit travel time $p=1$, vehicle driving cost rate $c=0.1$ and commission rate $R=0.5$. 



From the discussion of policy cycles in Section \ref{sec:pcycles}, one can infer that the value function in $\mathcal{OPT}$ might not be continuously differentiable since changing $\bm{b}^C$ might affect the cycles that get positive driver mass and hence the rate of generating profit, and this can happen in discrete points. We can observe this in Fig. \ref{fig:exhaustive}, where there are lines of non-differentiable points. At these points, the platform profit can be improved by varying the demand $\bm{b}^C$ revealed to CVs following a specific direction, but the direction cannot be predicted by taking partial derivatives along the horizontal and vertical directions at the points. Also, the surface of platform profit is not convex, with separated locally optimal structures at multiple places (e.g. $b_1^C=2$, $b_2^C \in [1.5,2.6]$).

We also observe from this example and in other extensive experiments on more two-region networks that the platform profit is piecewise linear. This is inline with the discussion on how changes in the demand revealed to CVs affect the equilibrium by changing the waiting times and thus the profitability of cycles. This might create challenges to traditional gradient descent optimization algorithms.

\subsection{Gradient-Descent Algorithm}

First note that for solving the bilevel optimization $\mathcal{OPT}$, the goal can be simplified into finding the optimal demand $\bm{b}^{C}$ revealed to CVs since following that, the platform schedules AVs to serve optimally the remaining demand $\bm{b}-\bm{b}^{C}$. Hence, we propose a gradient-descent algorithm Alg.~\ref{alg:gd}, updating $\bm{b}^{C}$ in the direction of gradient to find the maximum platform profit.

\begin{algorithm}[H]
\caption{Gradient-descent for computing $\{b_i^{C\star}\}$} \label{alg:gd}
\begin{algorithmic}[1]
\State Initialize the demand revealed to CVs $\bm{b}^C=\bm{b}^C_0$, and compute the corresponding platform profit $r_{old}$; 


\State{Select step size $s$ and termination parameter $\epsilon$;}
\State{Set $t=0$;} 
\While{$t< 100$}
    \For{ each region $i = 1:L$}        
        \State{Calculate the partial derivative of platform profit w.r.t. $b_i^{C}$ numerically, recorded as $d_i$;} 
        
         
    \EndFor

        
    \State{Update $b_i^C = b_i^C + s \cdot d_i  / ||\bm{d}||_1$, $\forall i$;}
   

    \State{Compute the platform profit $r_{new}$ by solving problem \eqref{equ:bilevel} and \eqref{equ:lowerlevel} with the updated $\{b_i^C\}$;}
    

    \If{$r_{new} - r_{old} \leq \epsilon$}
    \State{\textbf{break};}
    \EndIf
    \State{$r_{old} = r_{new}$;}
    \State{$t = t + 1$;}
    
\EndWhile

\State{\textbf{return} $\{b^C_i\}$}
\end{algorithmic}
\end{algorithm}

However, the platform profit given by \eqref{equ:obj2} does not exhibit convexity and is not differentiable everywhere across the entire domain of $\bm{b}^{C}$.
Thus, since we numerically compute the derivatives, Alg.~\ref{alg:gd} might fail to produce the correct `ascend' direction, especially at non-differentiable points. Even if the algorithm converges, it may terminate at a local optimum instead. From numerical experiments, we observe that the performance of the algorithm depends significantly on the initial point $\bm{b}^C_0$. To deal with the aforementioned issues, we propose to start from initial points that already generate high profit (e.g., the solution derived by the AV-first policy), or select multiple initial points and take the best result, to improve robustness. This is demonstrated to perform well in our simulation.

\subsection{Bundle Method Algorithm}

As has been discussed, the platform profit \eqref{equ:obj2} is not differentiable everywhere with respect to $\bm{b}^C$. A typical technique to deal with convex non-differentiable problems is the bundle method. Although our problem is not convex, the bundle method can provide a near-optimal solution in our simulation. In this subsection, we first introduce how the bundle method works when minimizing a convex function, then propose the algorithm to solve our mixed AV-CV problem \eqref{equ:bilevel}.

\begin{definition}[Moreau-Yosida Regularization]
Let $f: \mathbb{R}^n \to \mathbb{R}$ be a proper convex function. The Moreau-Yosida regularization for a fixed $\mu>0$ is given by
\begin{equation*}
F(x) = \min_{y\in \mathbb{R}^n}\{f(y)+\frac{\mu}{2}\|y-x\|^2\}.
\end{equation*}
The minimizer of the regularization is the proximal point
\begin{equation*}
p(x) = \arg \min_{y\in \mathbb{R}^n}\{f(y)+\frac{\mu}{2}\|y-x\|^2\}.
\end{equation*}
\end{definition}

The Moreau-Yosida regularization makes the tradeoff between minimizing $f$ and being in the neighborhood centered by $x$, where the radius of the neighborhood is controlled by parameter $\mu$. Properties such as differentiability and convexity of the regularization $F$ have been widely analyzed in textbooks such as \citealt{nonsmooth}. We know that if $x^* \in \arg\min f$, then $x^* \in \arg\min F$. Therefore, testing $x^* \in \arg\min F$ can be the stopping criterion of the algorithm searching for the minimizer of $f$.




Note that the information from the past iterations can be used to approximate the function $f$. Consider the bundle of information $\{y^i,f(y^i),\partial f(y^i)\}_{i=1}^l$. A piecewise linear approximation of $f$ can be constructed by the following called cutting plane model
\begin{equation}
\hat{f}_{l}(x):= \max_{i=1,...,l}\{f(y^i)+\left \langle s^i,x-y^i \right \rangle\}.
\end{equation}
By the convexity of $f$, $f(x) \ge f(y^i)+\left \langle s^i,x-y^i \right \rangle $ for all $x\in \mathbb{R}^n$ and $i=1,...,l$. Hence, this piecewise linear approximation is a lower bound of the original function, i.e., $\hat{f}_l(x)\le f(x)$ for all $x \in \mathbb{R}^n$. Also, the approximation gets closer to the original function as the bundle size grows, i.e., $\hat{f}_{l+1}(x) \ge \hat{f}_l(x)$ for all $x \in \mathbb{R}^n$.

The bundle method uses the sequence of points in the past iterations to construct the bundle, and introduces a concept called the bundle center to record the best point discovered so far. Making use of the bundle to build approximation $\hat{f}$, it computes the next iterate by finding the proximal point of the Moreau-Yosida regularization of $\hat{f}$ at the current center $\hat{x}$. The algorithm terminates if $\hat{f}$ is a good approximation of $f$ and the current center minimizes $F$. Notice that the objective of our model is to maximize the platform profit, which is a function of the demand revealed to CVs. We denote the platform profit as the function $r$, and modify the bundle method algorithm to solve this maximization problem as follows.


\begin{algorithm}[H]
\caption{Bundle Method} \label{alg:bundle}
\begin{algorithmic}[1]
\Statex \textbf{Step 1}. Initialize the bundle center $\hat{x}^0$ and proximal point $y^0 = \hat{x}^0$. Compute the corresponding platform profit $r(\hat{x}^0)$ and subgradient $s^0 \in \partial r(\hat{x}^0)$. Initialize the approximate objective $\hat{r}_0(y) = r(y^0) + \left \langle s^0,y-y^0\right \rangle$. Set termination parameter $\Bar{\delta}$ and center update parameter $m$. Initialize iteration time $k=0$.
\Statex \textbf{Step 2}. Compute proximal point
\begin{equation} y^{k+1}=\arg\min_{y} -\hat{r}_k(y)+\frac{\mu_k}{2}\|y-\hat{x}^k\|^2.
\end{equation}
\Statex \textbf{Step 3}. If $\delta_k:=-r(\hat{x}^k)-(-\hat{r}_k(y^{k+1})+\frac{\mu_k}{2}\|y^{k+1}-\hat{x}^k\|^2)  < \Bar{\delta}$, STOP.
\Statex \textbf{Step 4}. Compute platform profit $r(y^{k+1})$ and a subgradient $s^{k+1}\in \partial r(y^{k+1})$.
\Statex \textbf{Step 5}. If $r(y^{k+1})-r(\hat{x}^k)\ge m $, update the bundle center $\hat{x}^{k+1}=y^{k+1}$. Else, keep the current bundle center $\hat{x}^{k+1}=\hat{x}^k$.
\Statex \textbf{Step 6}. Update the approximate objective
\begin{equation*}
\hat{r}_{k+1}(y):= \min\{\hat{r}_k(y),r(y^{k+1})+\left \langle s^{k+1},y-y^{k+1} \right \rangle \}.
\end{equation*}
\Statex \textbf{Step 7}. Set $k=k+1$ and go to \textbf{Step 2}.
\end{algorithmic}
\end{algorithm}

Notice that the bundle method pursues to construct a piecewise linear approximation of the original function. By extensive numerical experiments, we find in many network settings, the platform profit is itself a piecewise linear function. Also, the termination of the bundle method is not affected by non-differentiable points, which seem to be abundant in large-scale network settings. Hence, we expect the bundle method to converge to a point with good performance, better than gradient-descent.


However, since the bundle method was originally designed to solve convex non-differentiable problems while our $\mathcal{OPT}$ problem is non-convex, several issues may arise. First, the bundle method may get trapped in local optima, failing to converge to global optima. Also, the function approximation built by subgradients may not be informative, resulting in the method converging slowly or circulating without convergence. We note that the converging process highly depends on the value of parameter $\mu$. However, methods to automatically tune $\mu$ cannot work effectively in our non-convex problem. To deal with these issues, similar to the gradient-descent algorithm, we implement the method using multiple initial points and use the best result.


\subsection{Genetic Algorithm}

The inherent complexity of non-convex and non-smooth problems poses a significant challenge for conventional optimization algorithms due to their tendency to converge prematurely at local optima. To deal with that, we adopt a heuristic optimization approach known as the genetic algorithm that can explore a broader solution space and circumvent the limitations often encountered by classical optimization techniques. This method demonstrates efficacy in tackling a variety of optimization problems that are not well suited for standard optimization algorithms. In the context of our specific mixed-fleet problem $\mathcal{OPT}$, the genetic algorithm proceeds as detailed below:

\begin{enumerate}

\item \textbf{Initialization:} 
initialize a group of $K$ chromosomes, labeled as $\{\bm{b}^{C,1}, \bm{b}^{C,2}, \dots, \bm{b}^{C,K}\}$, with each representing a feasible solution to our mixed-fleet optimization problem $\mathcal{OPT}$.


\item \textbf{Selection:} select two chromosomes as parents according to selection probabilities, which are determined by the order of the chromosomes concerning a fitness metric. We adopt the platform profit \eqref{equ:obj2} as the fitness metric and rank all chromosomes from the best to the worst according to their fitness values. Then, each chromosome is allocated a selection probability $p_\text{S}(r)$, where $r$ denotes its order.
By adopting the normalized geometric ranking scheme, the definition of $p_\text{S} (r)$ is given by:
\begin{align}\label{equ:prob}
p_\text{S}(r)=\frac{q}{1-(1-q)^{K}}(1-q)^{r-1},
\end{align} 
where $q$ is pre-determined based on the selection guidelines.

\item \textbf{Generation:} perform similar procedures as in biology to create an offspring from the selected parents for the next generation, such as crossover operation and mutation operation. Specifically, a uniform crossover operation is implemented with a crossover possibility $p_\text{C}$ to generate offspring by swapping genes (i.e., the elements $b_i^{C,k}$ in $\bm{b}^{C,k}$ for region $i$) from the parents randomly. Furthermore, to avoid being trapped in local optima, a mutation operation is implemented with a mutation possibility $p_\text{M}$ to randomly select a gene from the chromosome and replace it with randomly generated new genes. Note that, if the generated new chromosome fails to exhibit higher profit than its parents, the process of crossover or mutation will be repeated until a higher platform profit is attained. 

A pair of parents is selected to generate one offspring. At the end of the entire generation process, the entire population of chromosomes is replaced with a new set of generated chromosomes and the population size is kept constant. The process of selection and generation is repeated until the pre-specified maximum number of generations $T_{max}$ is reached or no further improvement is achieved within recent generations. The proposed algorithm is outlined as Alg.~\ref{alg:genetic}. 

\end{enumerate}

\begin{algorithm}[H]
\caption{Genetic Algorithm-based Scheme} \label{alg:genetic}
\begin{algorithmic}[1]
\Statex \textbf{Step 1}. Initialize a group of feasible chromosomes $\{\bm{b}^{C,k}\}$ with population size $K=10$. Set maximal iterations $T_{max}=100$ and $t=1$.
\Statex \textbf{Step 2}. Evaluate the fitness value of each chromosome $\bm{b}^{C,k}$ and rank them according to their values. Compute their selection probabilities using \eqref{equ:prob} with $q=0.1$.
\Statex \textbf{Step 3}. Select two chromosomes as parents according to the selection probabilities.
\Statex \textbf{Step 4}. Given the selected parents, generate an offspring for the next generation with the crossover operator of probability $p_\text{C}=0.5$ and mutation operator of probability $p_\text{M}=0.6$. If the offspring has a fitness value less than its parents, then repeat Step 4.
\Statex \textbf{Step 5}. Repeat Steps 3 and 4 until $K$ offsprings are generated to form a new population, and update iteration times $t=t+1$.
\Statex \textbf{Step 6}. If $t<T_{max}$ and there is an improvement compared with $10$ generations ago, go to Step 2, else terminate.
\end{algorithmic}
\end{algorithm}

In our simulation, in Alg.~\ref{alg:genetic}, a large population size $K$ can be set to deal with the irregular shape of the platform profit function against $\{b_i^{C}\}$, and the complexity of the algorithm increases linearly in $K$.

\section{Numerical Results} \label{sec:numerical}

As discussed in the last section, both gradient-descent and bundle method algorithms are initially designed to solve convex problems with a single optimum. Their performance in solving our non-convex and non-smooth problem $\mathcal{OPT}$ cannot be guaranteed due to their tendency to converge prematurely at local optima. Meanwhile, the genetic algorithm is a heuristic algorithm that involves a succession of random operations. Thus, to evaluate the performance of our proposed algorithms, in this section, we first conduct experiments involving the basic two-region network, comparing the algorithms' results with the analytically derived optimal solution. Further, we perform extensive numerical simulations for a broader scope of grid networks, for which the optimal solution can only be obtained by (time-prohibitive) exhaustive search. We compare our three proposed algorithms in terms of the profit values of the local optima they discover and their running time.


\subsection{Two-Region Networks}\label{sec:sm}

\begin{figure}
    \centering 
    \includegraphics[width=8cm]{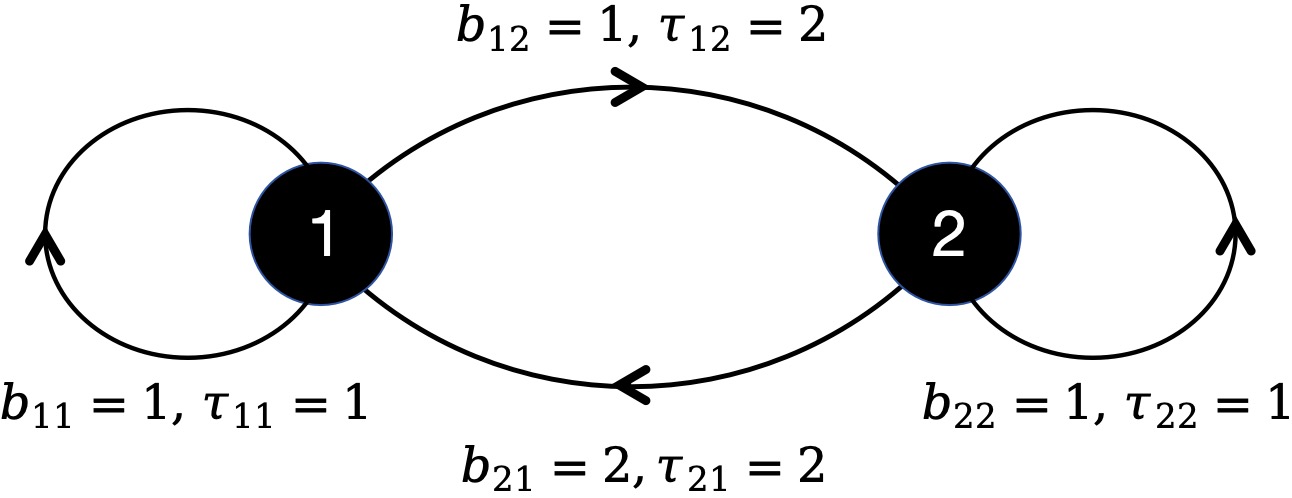}
    \caption{A two-region network with four customer routes: two within-region routes $1\to 1$, $2\to 2$ and two cross-region routes $1\to 2$, $2\to 1$. The demand rates and travel times are specified as $\{b_{11}=1, b_{12}=1, b_{21}=2, b_{22}=1\}$ and $\{\tau_{11}=1, \tau_{12}=2, \tau_{21}=2, \tau_{22}=1\}$ for each route. We assume the cross-region demands are imbalanced (i.e., $b_{21} > b_{12}$) and the traveling time of the cross-region route is greater than the within-region route.}
    \label{fig:figureL2}
\end{figure}

    


Our goal is to evaluate the performance of our three proposed algorithms with different values of AV fleet size $M$ in the two-region network in Fig.~\ref{fig:figureL2}, under two different levels of CV fleet size: low CV supply $N=5$ and high CV supply $N=10$. We refer to $N=10$ as the high supply because $b_{11} \tau_{11}+b_{21} \tau_{12}+b_{21}\tau_{21}+b_{22} \tau_{22}=10$ is the minimum number of vehicles required to serve all the demand in the two-region network if all vehicles are fully controlled\footnote{We use $b_{21} \tau_{12}$ instead of $b_{12} \tau_{12}$ in the above calculation because except the vehicles always busy with transporting customers, an additional rate $(b_{21}-b_{12})\tau_{12}$ of repositioning vehicles to region 2 is required for fully serving the remaining demand from region 2 to region 1.}. Moreover, as discussed in Section \ref{sec:heu}, given the considerable impact of the initial point on the performance of the gradient-descent and bundle method algorithms, we implement the two algorithms with multiple initial points (the four extreme points of the feasible space $[0,b_1] \times [0,b_2]$) and choose the best result in the two-region network.

\begin{figure}
    \centering 
    \includegraphics[width=10cm]{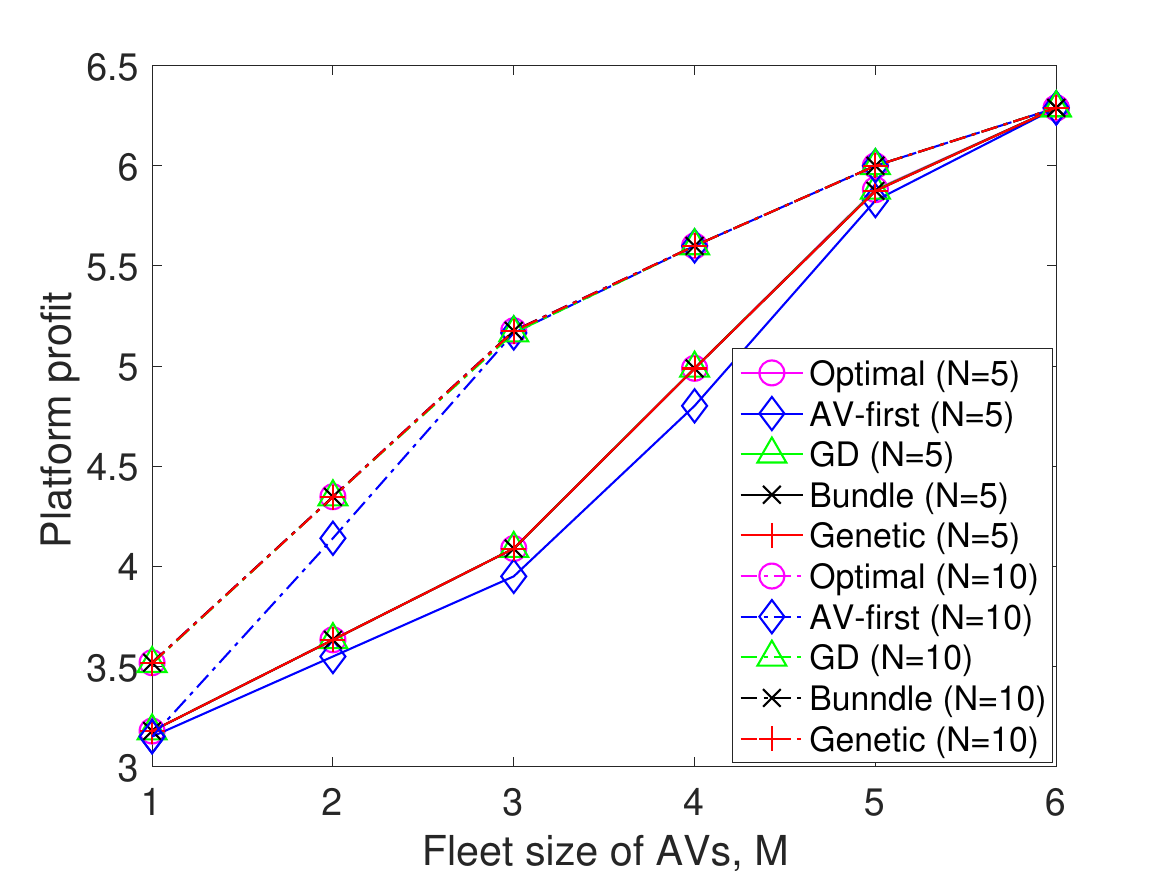}
    \caption{Compare the analytically derived optimal platform profit against the results obtained by the AV-first policy, gradient-descent, bundle method and genetic algorithms. Test the performance for various AV fleet sizes $M$, across two levels of CV supply (low CV supply $N=5$ and high CV supply $N=10$).\\
    \textbf{Observations}: 1. All three proposed algorithms attain near-optimal performance. 2. The maximum performance loss of AV-first is $10\%$ for this simulation (i.e., when $M=1$ and $N=10$). 3. For a given fleet size of AVs, a higher platform profit is achieved with a larger fleet size of CVs.
    }
    \label{fig:l2}
\end{figure}


In Fig.~\ref{fig:l2}, we compare the platform profit under the three proposed algorithms, AV-first policy, and the optimal solution. This comparison is made concerning the AV fleet size $M$ and two distinct levels of CV supply. We notice that all three proposed algorithms outperform the AV-first policy, delivering results close to the optimum. Even in the most challenging setting, where the AV-first policy exhibits a performance loss of up to $10\%$ (i.e., when $M=1$ and $N=10$), our proposed algorithms demonstrate a maximum performance drop of merely $0.14\%$ (by gradient-descent) compared to the optimal solution.

Moreover, there are some insights we can learn from Fig.~\ref{fig:l2}. First, given the AV fleet size $M$, a higher platform profit can be obtained by a higher CV supply, validating Lemma \ref{lem:monotone}). Also, as anticipated, for a given CV fleet size $N$, the platform profit consistently increases in the AV fleet size $M$.

\subsection{Larger Grids}

For larger networks with more than two regions, it is difficult to analytically derive the optimal solution of $\mathcal{OPT}$. However, it is essential to test the time complexity and the performance of the algorithms in larger-scale networks. Hence, we conduct experiments in the larger-scale grid networks, including $2\times2$, $3\times3$ and $4\times4$ grids, establishing the AV-first policy as the baseline for comparison due to its simplicity and ease of analysis. To define the grid networks, we assume that the regions are nodes evenly located in a two-dimensional Manhattan space. For example, in the $2\times2$ grid, the four regions are positioned at coordinates $(0,0)$, $(0,1)$, $(1,0)$, and $(1,1)$. The traveling time between any pair of regions is determined by their Manhattan distance, while the demand between them is uniformly selected from the set $\{0,1,2\}$. We maintain the same customer payment per unit travel time of $p = 1$ and vehicle driving cost rate of $c = 0.1$ as in the two-region network, and set a larger commission rate of $R = 0.7$ to expand the potential improvement of our algorithms over the simple AV-first policy. Furthermore, we assume the fleet sizes of AVs and CVs to be $M=8$ and $N=16$ in $2\times2$ grids, $M=40$ and $N=80$ in $3\times3$ grids, and $M=200$ and $N=400$ in $4\times4$ grids, respectively. This scaling is necessary as the required number of vehicles to fulfill all the demand increases super-linearly with the network size.

\begin{table*}[htbp]  
\TABLE
    {\label{tab:compare}}
    {\begin{tabular}{c|c|cc|cc|cc}
     & \multirow{2}{*}{Initial points} & \multicolumn{2}{c|}{Gradient-descent} & \multicolumn{2}{c|}{Bundle method} & \multicolumn{2}{c}{Genetic}\\
  &  & Extra profit& Time&Extra profit & Time&Extra profit & Time\\   \hline  
  
 \multirow{4}{*}{ $2\times 2$ grid} & \multirow{2}{*}{ from AV-first }   &  \multirow{2}{*}{$3.01\%$}& \multirow{2}{*}{1.07} &\multirow{2}{*}{$3.26\%$}& \multirow{2}{*}{4.16}  &\multirow{4}{*}{$4.11\%$}&\multirow{4}{*}{21.67}\\
  & & & & & & & \\ 

  &\multirow{2}{*}{ from $\mathcal{B}_0$} &  \multirow{2}{*}{$2.24\%$}& \multirow{2}{*}{19.49}&\multirow{2}{*}{$3.91\%$}& \multirow{2}{*}{15.44}& & \\ 

 & & & & & & & \\ 
 \hline 

  \multirow{4}{*}{ $3\times 3$ grid}   & \multirow{2}{*}{ from AV-first }   &  \multirow{2}{*}{$1.44\%$}& \multirow{2}{*}{3.39} &\multirow{2}{*}{$0.73\%$}& \multirow{2}{*}{10.91}  &\multirow{4}{*}{$3.86\%$}&\multirow{4}{*}{32.57}\\
 & & & & & & & \\ 

& \multirow{2}{*}{ from $\mathcal{B}_0$ }&  \multirow{2}{*}{$1.31\%$}& \multirow{2}{*}{30.65}&\multirow{2}{*}{$2.12\%$}& \multirow{2}{*}{50.96}& & \\ 

& & & & & & & \\    
   \hline 
  
   \multirow{4}{*}{ $4\times 4$ grid}     &\multirow{2}{*}{ from AV-first }   &  \multirow{2}{*}{$0.56\%$}& \multirow{2}{*}{2.74} &\multirow{2}{*}{$0.22\%$}& \multirow{2}{*}{66.62}  &\multirow{4}{*}{$4.78\%$}&\multirow{4}{*}{47.87}\\
& & & & & & & \\ 

 & \multirow{2}{*}{ from $\mathcal{B}_0$} & \multirow{2}{*}{$3.06\%$}& \multirow{2}{*}{76.74}&\multirow{2}{*}{$0.55\%$}& \multirow{2}{*}{123.79}& & \\ 
& & & & & & & \\ 
    \end{tabular}

   } {We compare the running times (in minutes) and the \% extra profit obtained by the algorithms compared to AV-first  for the case of $2\times2$, $3\times3$ and $4\times4$ grids (no optimum solution available), for different initial points: the AV-first solution and the set of uniformly selected points $\mathcal{B}_0$. Each data point is averaged across $10$ randomly generated networks of each size and when using $\mathcal{B}_0$ we keep the best result.
 We observe that all three algorithms outperform the AV-first, the genetic scoring the best. For the gradient-descent and bundle method, the initial points can make a big difference.}
\end{table*}

In Table \ref{tab:compare}, we present the relative increase in platform profit compared with the AV-first and running time for the gradient-descent, bundle method and genetic algorithms in the $2\times2$, $3\times3$ and $4\times4$ grid setups. The results are averaged over $10$ simulations, each with a randomly generated demand matrix. We execute gradient-descent and bundle method algorithms with two settings of initial points $\bm{b}^C_0$: one uses as $\bm{b}^C_0$ the solution derived by the AV-first policy as the initial point, and the other uses a set $\mathcal{B}_0$ of five evenly chosen points in the feasible space $0\leq \bm{b}^C_0\leq \bm{b}$, i.e., $\mathcal{B}_0=\{\bm{0},\frac{1}{4}\bm{b},\frac{1}{2}\bm{b},\frac{3}{4}\bm{b},\bm{b}\}$. In the latter case, the best among the five results are selected for further profit comparison.

It can be observed from Table \ref{tab:compare} that all of our proposed algorithms outperform the AV-first policy in all network cases. The genetic algorithm performs consistently better, and the bundle method performs in most cases better than the gradient-descent. It is interesting to observe that the running time of the genetic algorithm is less influenced by the networks' scale compared to the other two algorithms. This might be explained primarily by the expedited convergence of the genetic algorithm in large networks. Recall that the genetic algorithm terminates if no improvement is achieved within the recent generations/iterations. When the dimensionality of the optimization problem is larger, this might happen easily due to the failure of the exploration for finding the correct improvement direction among all the potential directions. Another interesting observation is that the gradient-descent algorithm converges faster in the $4\times 4$ grid than in the $3\times 3$ grid when initiated from the solution of the AV-first. This is mainly because the running time of gradient-descent not only depends on the initial point and dimensionality of the problem (or network size), but also on the degree of the non-convexity of the problem (number of local optima). The structure of $\mathcal{OPT}$ for a larger grid has more local optima, which makes gradient-descent easier to get trapped, and hence terminate earlier than running for a smaller grid.

\begin{table*}[htbp] 
\TABLE
{\label{tab:diffinitial}}
{\begin{tabular}{c|ccccccccccccc} 
\hline
\hline
\diagbox{Algorithms}{Initial points}
& AV-first  & $\bm{0}$ & $\frac{1}{4}\bm{b}$ & $\frac{1}{2}\bm{b}$ & $\frac{3}{4}\bm{b}$ & $\bm{b}$ & Std\\
\hline

Gradient-descent 
& 13.89& 10.55 &12.70  & 13.52  & 12.60 & 10.72 & 1.31 \\

Bundle method 
& 14.08& 14.07 & 14.19 & 14.00 & 14.32 & 14.07 & 0.13 \\

Genetic 
& \multicolumn{6}{c}{14.77} & -\\
\hline
\hline
\end{tabular}
}{The platform profits computed by our three algorithms for the $2\times2$ grid network with demand  $\bm{b}=\begin{pmatrix} 
0 & 2 & 1 &2 \\
0 & 0 & 1 &2 \\
1 & 2 & 0 &2 \\
0 & 2 & 2 &0
\end{pmatrix}$ using different initial points. We observe that the performance of the gradient-descent and bundle method is significantly influenced by the choice of the initial point. Also, compared to the bundle method, the gradient-descent is easier to get trapped in sub-optimal points, leading to a higher standard deviation of the profits for different initial points (last column).}
\end{table*}

It is not surprising that if initiated from the solution derived by the AV-first policy, both the gradient-descent and bundle method algorithms can attain better performance than AV-first. Moreover, it is noteworthy that the two algorithms also demonstrate better performance than AV-first when initiated from the five evenly selected points, and sometimes even outperform the AV-first initialization. To better understand the performance of gradient-descent and bundle method, we take a closer look at the profit obtained using different initial points. Table \ref{tab:diffinitial} illustrates the computed results for a $2\times2$ grid (see Appendix \ref{app:2grid} for the results of all $10$ instances). It can be computed that the platform profit of the scheduling provided by the AV-first is $12.85$. Table \ref{tab:diffinitial} shows that when implementing the gradient-descent algorithm, many initial points yield a profit even lower than the profit of AV-first, resulting in a large variance of the algorithm. Nonetheless, by taking the best result among the multiple initial points, gradient-descent can achieve a higher profit than the AV-first.

Additionally, besides the average results shown in Table \ref{tab:compare}, we also observe that the improvement in profit proposed by our algorithms compared with the AV-first never exceeds $10\%$ in all the simulations, aligning with Conjecture \ref{conj:gap} that the performance loss of AV-first is no larger than $20\%$.



\subsection{Insights on Marginal Value of AVs} 

It is interesting to observe from Fig.~\ref{fig:l2} that in the case of low CV supply (i.e., $N=5$) the marginal value of adding an AV is not monotonically decreasing as expected, i.e., the curve of platform profit becomes steeper at $M=3$. To understand that, it is crucial to recognize that the marginal value is influenced not only by the left unserved demand, which typically becomes less profitable as the number of vehicles increases, but also by the inefficient queueing behavior of CVs. For example, when $N=5$ and the AV fleet size $M=2$ or $3$, the most profitable demand (those can be served without vehicles repositioning) is always fully served, by first allocating to AVs and then the remaining to CVs. Meanwhile, there are excess CVs that tend to queue in region 1 rather than reposition to region 2 due to the costly repositioning process. Thus, as $M$ increases from $2$ to $3$, one unit of CV previously serving the most profitable demand is displaced by AVs and stays idle in the queue, thereby increasing the profit since AVs can earn more than CVs to the platform for serving the same demand. However, with a further increase in AV fleet size from $3$ to $4$, more CVs are displaced by AVs, resulting in the CV queue long enough to encourage CVs repositioning. Then, the repositioning CVs leads to an additional profit increase by serving customers in region 2.

\begin{figure}
    \centering 
    \includegraphics[width=10cm]{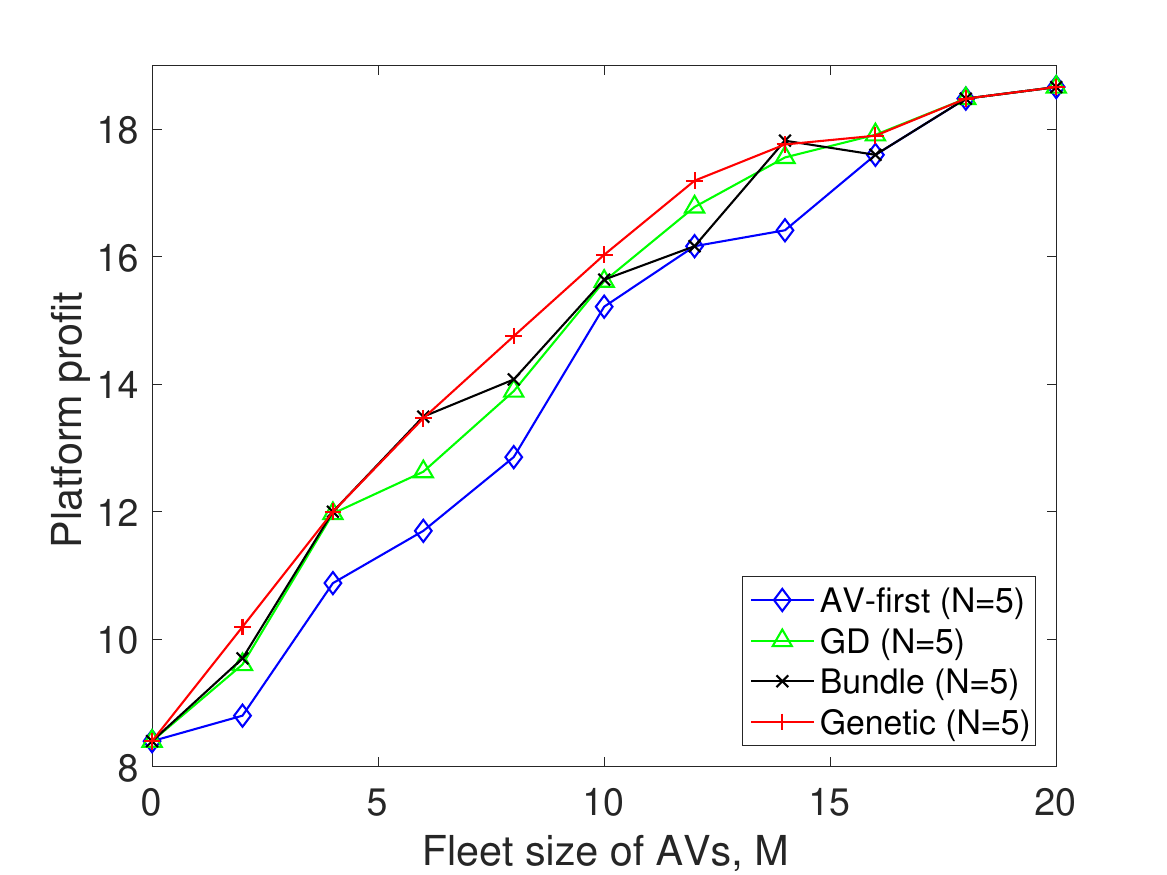}
    \caption{Computed platform profits of the $2\times2$ grid network in Table \ref{tab:diffinitial} against AV fleet size $M$.\\
    \textbf{Observation}: 1. The marginal value of adding AVs is not monotonically decreasing, e.g. at $M=2$ and $M=8$. 2. The marginal values depend a lot on which algorithm is implemented.}
    \label{fig:marginal}
\end{figure}


One can expect that in larger networks, the curve of platform profit might become steeper (with increasing marginal value) more than once at multiple AV fleet sizes. This is because as the fleet size of AVs increases, queues are gradually built up in multiple regions. 


We depict Fig.~\ref{fig:marginal} to show the platform profit against the AV fleet size $M$ in the instance of $2\times 2$ grid given in Table \ref{tab:diffinitial}. Notice that the time complexity for computing the optimal solution to $\mathcal{OPT}$ using exhaustive search is prohibitive in the $4$-region network. Instead, we implement the AV-first policy and the three proposed algorithms for different values of $M$. We observe that there exist multiple steep increases of the slope under the AV-first, gradient-descent and bundle method algorithms, e.g., at $M=2$ and $M=8$,
which are related to the queue behaviours of CVs in region 2 and region 4, respectively. Another inspection from the figure is that the marginal value depends on which algorithm is executed. For example, the bundle method exhibits an increasing marginal value at $M=12$ while the marginal value computed by gradient-descent and genetic algorithm keep decreasing at this point. The curve for the genetic algorithm is the most smooth with only one increase of the slope observed at $M=16$.

Such an observation on the marginal value of AVs suggests a more sophisticated analysis of introducing AVs when considering the purchase costs of AVs. In the following section, we will study how to extend our model to optimize the platform's decisions on the procurement quantity of AV.

\section{Extension to Endogenous Supply}\label{sec:endo}

In practical implementation, the platform's strategic integration of AVs for ride-hailing services necessitates an assessment of the purchase costs of AVs to optimize the procurement quantity. Furthermore, with the introduction of AVs, incumbent CV drivers may opt out as a result of diminishing anticipated earnings. This section delves into an exploration of our analysis, broadening the scope to consider the endogenous supply of AVs and CVs. Here, the platform confronts decisions regarding the procurement quantity of AV and drivers opt into or out of platform engagement depending on whether or not their expected earnings working for the platform exceed their opportunity costs. Here we refer to a driver's earning from outside job options as his opportunity cost.

\subsection{Endogenous Supply of AVs}
Our initial focus lies on the endogenous supply of AVs while maintaining a constant number of participating CV drivers. Let $I$ denote the AV purchase cost amortized over the expected operational lifespan of an AV. Then, to maximize the total platform profit that takes into account the purchase cost of AVs, we modify our basic model to construct a bi-level optimization problem, where at the upper level the platform determines not only the AV strategy $\bm{x}^A$ and the revealed demand $\bm{b}^C$ to CVs, but also the quantity $M$ of AVs to purchase. To simplify the problem, we first notice that acquiring a surplus of AVs that remain inactive in parking locations is never optimal for the platform. Thus, the designated AV fleet size $M$ aligns precisely with the aggregate of active AVs, i.e., $M =\sum_{i=1}^L\sum_{\alpha=1}^L t_{i\alpha}x_{i\alpha}^A$. Leveraging this insight, we reformulate the overall platform profit maximization problem as follows:

\begin{subequations}\label{equ:all}
\begin{align}
\mathcal{OPT}_{\mathcal{I}}:\quad\max_{\bm{x}^A,\bm{b}^C} \quad & \sum_{i=1}^L\sum_{\alpha=1}^L r_{i\alpha}^A x_{i\alpha}^A +\pi(\bm{b}^C)-I\sum_{i=1}^L\sum_{\alpha=1}^L \tau_{i\alpha}^{dr} x_{i\alpha}^A\label{equ:obj3}\\
\text{s.t.} \quad & \eqref{equ:demand},\eqref{equ:avbalance},\eqref{equ:range},\nonumber
\end{align}
\end{subequations}
where there exists an additional term in the objective function \eqref{equ:obj3} indicating paying the cost of purchasing AVs. 

To address problem $\mathcal{OPT}_{\mathcal{I}}$ efficiently, we can modify the three proposed algorithms for solving $\mathcal{OPT}$ given in Section \ref{sec:heu}, substituting \eqref{equ:obj2} by \eqref{equ:obj3} in these algorithms.

To provide a simple solution to problem $\mathcal{OPT}_{\mathcal{I}}$, we define the AV-first policy for this extended problem as well.

\begin{definition}
    AV-first Policy for $\mathcal{OPT}_{\mathcal{I}}$: Without considering the existence of CVs, deploy the AVs by determining the AV strategy $\bm{x}^A$ to maximize platform profit, which includes AV purchase costs. If there is any residual demand, then reveal it to CVs.
\end{definition}

\subsection{Endogenous Supply of CVs}

Based on the new problem $\mathcal{OPT}_{\mathcal{I}}$, we further extend our analysis to encompass the endogenous supply of CVs. In this paradigm, CV drivers can decide whether or not to work for the platform depending on their expected earnings from the platform and outside job opportunities. We assume there is a continuum of drivers of mass $N_{max}$, with heterogeneous opportunity costs following a uniform distribution across the interval $[0,(1-R)p-c]$, where $(1-R)p-c$ is the maximum profit achievable by an individual CV driver involving in the platform. That is to say, the maximum expected earning $(1-R)p-c$ corresponds to all potential drivers $N_{max}$ working for the platform. If the expected earning is less than $(1-R)p-c$, say $U$, then a portion $U/((1-R)p-c)$ of total $N_{max}$ drivers will stay in the system.

For this extended setting, we can continue to derive a bi-level optimization problem where the upper level remains the same as in $\mathcal{OPT}_{\mathcal{I}}$. The lower level problem solves the CV equilibrium not only satisfying Definition \ref{def:CVE}, but also following that the expected earning of working in the platform cannot be smaller than the participated drivers' opportunity costs. We define the expected individual CV earning $u(N)=\sum_{i=1}^L\sum_{\alpha=1}^L r_{i\alpha}^C x_{i \alpha}^C/N$ as a function of the number $N$ of participating drivers, where $\{x_{i\alpha}^C\}$ is the CV equilibrium of fleet size $N$ (the solution of $\mathcal{CV}(\bm{b}^C)$ with $N$).

Because of the uniformly distributed opportunity cost in interval $[0,(1-R)p-c]$, the CV fleet size $N$ should satisfy $N/N_{max}=u(N)/((1-R)p-c)$, which can be further transformed into
\begin{equation*}
    u(N)=\frac{((1-R)p-c)N}{N_{max}}.
\end{equation*}
The RHS of the equation increases in $N$ while the earning of participation in the LHS exhibits a non-increasing trend in $N$. Therefore, we can employ the bisection method to compute the equilibrium fleet size $N$ and the corresponding equilibrium action rates.


To combine the bisection method and our proposed algorithms to obtain the solution, in the steps that need computation of the platform profit for a given $\bm{b}^C$ (e.g., line 9 in Alg.~\ref{alg:gd}), to solve the CV equilibrium, we execute the following procedure to determine the CV fleet size $N$. Initialize with the lower bound $N_1=0$ and the upper bound $N_2=N_{max}$, then:


\begin{enumerate}
    \item Given the revealed demand $\bm{b}^C$, compute the CV equilibrium with fixed fleet size $N=N_{mid}:=(N_1+N_2)/2$, i.e., substitute $N_{mid}$ into problem $\mathcal{CV}(\bm{b}^C)$ and solve.
    \item Update $N_1=N_{mid}$ if $u(N_{mid})>((1-R)p-c)N_{mid}/N_{max}$, otherwise update $N_2=N_{mid}$ if $u(N_{mid})<((1-R)p-c)N_{mid}/N_{max}$. Return to step 1 and repeat this procedure until $u(N_{mid})=((1-R)p-c)N_{mid}/N_{max}$.
\end{enumerate}


\subsection{Numerical Results}
In this subsection, we conduct an extensive experiment to demonstrate the effectiveness of the modified algorithms in endogenous supply scenarios by comparing their results with the analytically derived solution. Notice that a parallel analysis of two-region networks for the endogenous setting can be conducted similarly to the basic exogenous setting. This analysis provides the closed-form AV/CV action rates in the optimal solution as a function of the AV purchase cost $I$ and the mass of potential CV drivers $N_{max}$. The detailed derivation is given in Appendix \ref{app:endo}.

\begin{figure}
    \centering 
    \includegraphics[width=10cm]{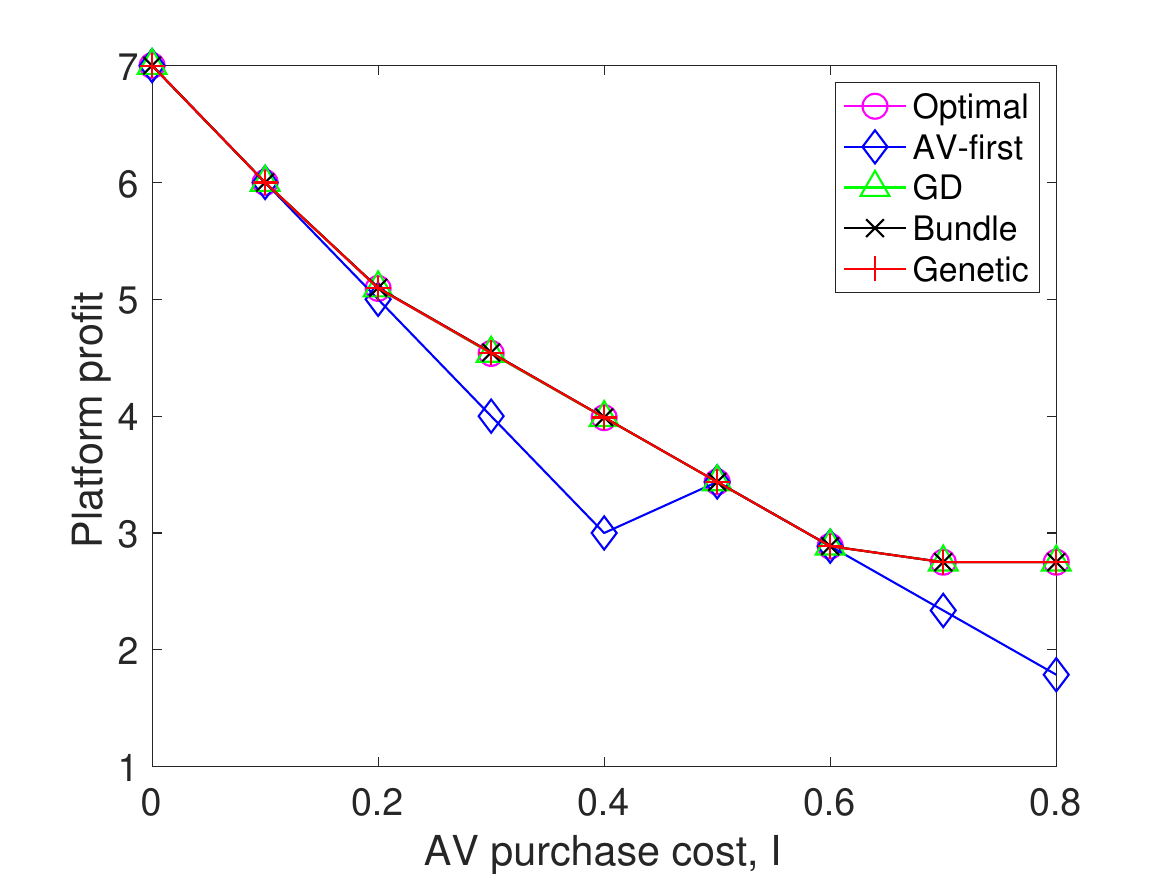}
    \caption{Platform profit of the network shown in Fig.~\ref{fig:figureL2} against the AV purchase cost $I$ computed by the modified three proposed algorithms, the AV-first policy, and the analytically derived optimal solution.\\
    \textbf{Observation}: 1. All three algorithms exhibit near-optimal performance. 2. The maximum performance loss of AV-first is $35.02\%$ at $I=0.8$, implying the $20\%$ upper bound of its performance loss does not hold for endogenous settings.}
    \label{fig:endo}
\end{figure}



We conduct experiments on the two-region network shown in Fig.~\ref{fig:figureL2} that was used in the previous exogenous scenario. We assume a pool size of $N_{max}=10$ potential CV drivers and the AV purchase cost $I$ ranging within interval $[0, I_{max} = p - c = 0.9]$. Notice that if the AV purchase cost exceeds $I_{max}$, the platform refrains from purchasing any AVs. 

Fig.~\ref{fig:endo} presents the platform profit against AV purchase cost $I$ derived by the three modified algorithms, the AV-first policy, and the analytically derived optimal solution. Notably, we can observe that all three proposed algorithms exhibit performance close to the optimal solution. Interestingly, the AV-first policy incurs a performance decline of up to $35.02\%$ at $I=0.8$, implying that the $20\%$ upper limit of AV-first performance degradation in two-region networks no longer holds in the endogenous setting.

\section{Conclusions}
In this paper, we formulate a novel bi-level optimization problem $\mathcal{OPT}$ to maximize the total profit of the ride-hailing platform generated by AVs and CVs for networks with arbitrary number of regions. At the upper level, the platform as the leader decides on the state-action rates of AVs and the demand revealed to CVs. Then, at the lower level, CVs as the follower respond by forming an equilibrium. The key idea used here is to adopt the mean-field game equilibrium analysis in \citealt{antonis} to formulate the computation of the CV equilibrium as a convex optimization problem. However, due to the implicit nature of the profit generated by CVs in the equilibrium and its non-monotonicity in the revealed demand, $\mathcal{OPT}$ is challenging to solve as a non-convex and non-smooth problem. 

To analyze $\mathcal{OPT}$, we first prove the `AV fully active' property that in the optimal solution, if there are CVs serving demand, then all AVs must have been used. Also, we show the straightforward AV-first policy is not optimal by giving an example where the platform prefers to offer the high-profit demand (demand that can be served without vehicle repositioning) first to CVs and use AVs to serve the demand that would otherwise not be served under AV-first. Further, we prove a $20\%$ upper limit for the performance degradation of the AV-first as compared to the optimum in two-region networks and conduct extensive numerical simulations to show the consistency of the limit for larger networks. Finally, we propose the gradient-descent, bundle method, and genetic algorithms to solve $\mathcal{OPT}$ efficiently. We show that all our proposed algorithms attain the optimum in two-region networks and exhibit non-negligible performance improvement compared to the AV-first in larger grid networks. Finally, we extend our analysis to the endogenous supply of AVs and CVs and modify our three algorithms to solve for the endogenous scenario.

An interesting direction of future research is to consider the admission control that allows the platform to assign empty vehicles to customer requests depending on the customers' destinations. In this way, the platform controls not only the aggregated demand $b_i^A$/$b_i^C$ in each region for AVs/CVs, but also the customer routing probability $q_{ij}$. This adds more flexibility in maximizing the platform profit through the control of the platform.

\bibliographystyle{nonumber}

\newpage

\begin{APPENDICES}

\section{Proof of Proposition \ref{prop:NE} } \label{app:cvx}
We first prove the ``only if" direction. Let $(\bm{w}^C = \{w_{i}^C\},\bm{x}^C = \{x_{i\alpha}^C\})$ be an equilibrium in the pure CV system with a given demand $\bm{b}^C$. Next, we will show it is the solution of $\mathcal{CV}(\bm{b}^C)$. 

According to condition i) in Definition \ref{def:CVE} of CV equilibrium, $\bm{x}^C$ is the solution to \eqref{equ:cvlp} and maximizes the CV profit $\sum_{i=1}^L \sum_{\alpha=1}^L r_{i\alpha}^C x_{i\alpha}^C$ given in the objective function \eqref{equ:cvprofit}. This implies it is also the maximizer
of $N\log\sum_{i=1}^L\sum_{\alpha=1}^L r_{i\alpha}^C x_{i \alpha}^C$ under the same constraints. The optimality conditions for this new problem are

a) Feasibility: 
\begin{align}
\sum_{j=1}^L (\sum_{k=1}^L x_{kj}^C) q_{ji} = \sum_{\alpha = 1}^L x_{i \alpha}^C, \quad \forall i,   \label{equ:fea1}
\end{align}

\begin{align}
\sum_{i=1}^L \sum_{\alpha=1}^L (\tau^{dr}_{i\alpha}+w_{\alpha}^C)x_{i\alpha}^C = N.   \label{equ:fea2}
\end{align}

b) First order conditions:
\begin{align}
N\frac{r_{i\alpha}^C}{\sum_{i=1}^L\sum_{\alpha=1}^L r_{i\alpha}^C x_{i \alpha}^C}-(\tau^{dr}_{i\alpha}+w_{\alpha}^C)\nu-\sum_{j=1}^L \lambda_j q_{\alpha j} +\lambda_i\le 0 \text{ with equality if } x_{i \alpha}^C>0, \quad \forall i,\alpha\label{equ:firstorder}
\end{align}
where $\lambda_i$ and $\nu$ are the dual variables of \eqref{equ:fea1} and \eqref{equ:fea2}, respectively.

First note that multiplying both sides of \eqref{equ:firstorder} with $x_{i\alpha}^C$, summing over $i$ and $\alpha$, and applying \eqref{equ:fea1} and \eqref{equ:fea2} yields $\nu=1$. Thus, the optimality conditions above can be rewritten as

a') Feasibility: 
\[\sum_{j=1}^L (\sum_{k=1}^L x_{kj}^C) q_{ji} = \sum_{\alpha = 1}^L x_{i \alpha}^C, \quad \forall i. \]

b') First order conditions:
\[N\frac{r_{i\alpha}^C}{\sum_{i=1}^L\sum_{\alpha=1}^L r_{i\alpha}^C x_{i \alpha}^C}-\tau^{dr}_{i\alpha}-w_{\alpha}^C-\sum_{j=1}^L \lambda_j q_{\alpha j} +\lambda_i\le 0 \text{ with equality if } x_{i \alpha}^C>0, \quad \forall i,\alpha\]

Then, by combining a') and b') with the conditions ii) and iii) in Definition \ref{def:CVE} of CV equilibrium (i.e., $\sum_{j=1}^L x_{ji}^C\le b_i^C,\forall i$ and $w_i^C(b_i^C-\sum_{j=1}^L x_{ji}^C)=0,\forall i$), we find they construct the optimality conditions for the problem $\mathcal{CV}(\bm{b}^C)$. Thus, $\bm{x}^C$ solves $\mathcal{CV}(\bm{b}^C)$ and $\bm{w}^C$ are the Lagrange multipliers of constraint \eqref{equ:cvdemand}.

To prove the ``if" direction, we also use the fact that the optimality conditions of problem $\mathcal{CV}(\bm{b}^C)$ for $(\bm{x}^C,\bm{w}^C)$ includes the conditions ii) and iii) given in Definition \ref{def:CVE} and the above conditions a'), b'). This further implies the solution of $\mathcal{CV}(\bm{b}^C)$ satisfies the optimality conditions of a), b) and thus condition i) in Definition \ref{def:CVE} also holds. Since $(\bm{x}^C,\bm{w}^C)$ satisfies all the three conditions given in Definition \ref{def:CVE}, $(\bm{x}^C,\bm{w}^C)$ is an equilibrium.

\section{Proof of Proposition \ref{pro:uniqueness} }\label{app:uniqueness}
\subsection{Property (1)}
As proved in \citealt{antonis}, the values of CV profit $\sum_{i=1}^L\sum_{\alpha=1}^Lr_{i\alpha}^C x_{i \alpha}^C$ and CV active mass $\sum_{i=1}^L\sum_{\alpha=1}^L \tau_{i\alpha}^{dr} x_{i \alpha}^C$ are the same for all equilibria, i.e., all solutions to \eqref{equ:lowerlevel}. Further, the profit CVs generate to the platform is also the same since it is given by
\begin{align}
\sum_{i=1}^L\sum_{\alpha=1}^L r_{i\alpha}^{C2P} x_{i \alpha}^C\hspace{-3pt}=\hspace{-3pt}\frac{R}{1\hspace{-3pt}-\hspace{-3pt}R}(\sum_{i=1}^L\sum_{\alpha=1}^L r_{i\alpha}^C x_{i \alpha}^C+c\sum_{i=1}^L\sum_{\alpha=1}^L \tau^{dr}_{i\alpha}x^C_{i \alpha}).  \label{equ:transfer}  
\end{align}

\subsection{Property (2)}
Proved in \citealt{antonis}

\subsection{Property (3)}

Based on \eqref{equ:transfer}, this is equivalent to proving the CV profit $\sum_{i=1}^L\sum_{\alpha=1}^Lr_{i\alpha}^C x_{i \alpha}^C$ and CV active mass $\sum_{i=1}^L\sum_{\alpha=1}^L \tau^{dr}_{i\alpha} x_{i \alpha}^C$ is non-decreasing with CV fleet size $N$. 

Consider two possible CV fleet sizes $N_1$ and $N_2$ with $N_1<N_2$. We use $\tilde{x}_{i \alpha}^C$ and $\hat{x}_{i \alpha}^C$ to denote the optimal solution of \eqref{equ:lowerlevel} with $N=N_1$ and $N=N_2$, respectively. First note that when $N$ increases from $N_1$ to $N_2$, the feasible region defined by \eqref{equ:cvdemand}-\eqref{equ:lezero} remains the same and thus $\tilde{x}_{i \alpha}^C$ and $\hat{x}_{i \alpha}^C$ are feasible for both problems with $N=N_1$ and $N=N_2$. 

As $N$ increases from $N_1$ to $N_2$, there are four possible cases for the changes of the logarithmic reward term $\log\sum_{i=1}^L\sum_{\alpha=1}^Lr_{i\alpha}^C x_{i \alpha}^C$ and the mass term $\sum_{i=1}^L\sum_{\alpha=1}^L \tau^{dr}_{i\alpha} x_{i \alpha}^C$ of \eqref{equ:objstage2}: 

i) the logarithmic reward term increases and the mass term decreases, i.e.,
\[\log\sum_{i=1}^L\sum_{\alpha=1}^Lr_{i\alpha}^C \tilde{x}_{i \alpha}^C< \log\sum_{i=1}^L\sum_{\alpha=1}^L r_{i\alpha}^C \hat{x}_{i \alpha}^C,\] 
\[\sum_{i=1}^L \sum_{\alpha=1}^L \tau^{dr}_{i\alpha} \tilde{x}_{i\alpha}^C > \sum_{i=1}^L\sum_{\alpha=1}^L \tau^{dr}_{i\alpha} \hat{x}_{i\alpha}^C,\]

ii) the logarithmic reward term decreases and the mass term increases, i.e.,
\[\log\sum_{i=1}^L\sum_{\alpha=1}^Lr_{i\alpha}^C \tilde{x}_{i \alpha}^C> \log\sum_{i=1}^L\sum_{\alpha=1}^L r_{i\alpha}^C \hat{x}_{i \alpha}^C,\] 
\[\sum_{i=1}^L \sum_{\alpha=1}^L \tau^{dr}_{i\alpha} \tilde{x}_{i\alpha}^C < \sum_{i=1}^L\sum_{\alpha=1}^L \tau^{dr}_{i\alpha} \hat{x}_{i\alpha}^C,\]

iii) both terms decrease, i.e.,
\[\log\sum_{i=1}^L\sum_{\alpha=1}^Lr_{i\alpha}^C \tilde{x}_{i \alpha}^C> \log\sum_{i=1}^L\sum_{\alpha=1}^L r_{i\alpha}^C \hat{x}_{i \alpha}^C,\] 
\[\sum_{i=1}^L \sum_{\alpha=1}^L \tau^{dr}_{i\alpha} \tilde{x}_{i\alpha}^C > \sum_{i=1}^L\sum_{\alpha=1}^L \tau^{dr}_{i\alpha} \hat{x}_{i\alpha}^C,\]

iv) both terms increase, i.e.,
\[\log\sum_{i=1}^L\sum_{\alpha=1}^Lr_{i\alpha}^C \tilde{x}_{i \alpha}^C< \log\sum_{i=1}^L\sum_{\alpha=1}^L r_{i\alpha}^C \hat{x}_{i \alpha}^C,\] 
\[\sum_{i=1}^L \sum_{\alpha=1}^L \tau^{dr}_{i\alpha} \tilde{x}_{i\alpha}^C < \sum_{i=1}^L\sum_{\alpha=1}^L \tau^{dr}_{i\alpha} \hat{x}_{i\alpha}^C.\]

Then, we show that case iv is the only possible case. First, case i is impossible, otherwise $\tilde{x}_{i \alpha}^C$ can never be the optimal solution to \eqref{equ:lowerlevel} with $N=N_1$. Similarly, case ii is also impossible, otherwise $\hat{x}_{i \alpha}^C$ can never be the optimal solution to \eqref{equ:lowerlevel} with $N=N_2$. As for case iii, we first note that we have
\[N_1 (\log\sum_{i=1}^L\sum_{\alpha=1}^L r_{i\alpha}^C \tilde{x}_{i \alpha}^C - \log\sum_{i=1}^L\sum_{\alpha=1}^Lr_{i\alpha}^C \hat{x}_{i \alpha}^C)< \]
\[N_2 (\log\sum_{i=1}^L\sum_{\alpha=1}^L r_{i\alpha}^C \tilde{x}_{i \alpha}^C - \log\sum_{i=1}^L\sum_{\alpha=1}^Lr_{i\alpha}^C \hat{x}_{i \alpha}^C)< \]
\[
\sum_{i=1}^L\sum_{\alpha=1}^L \tau^{dr}_{i\alpha} \tilde{x}_{i\alpha}^C- \sum_{i=1}^L \sum_{\alpha=1}^L \tau^{dr}_{i\alpha} \hat{x}_{i\alpha}^C. 
\]
where the first inequality is because $N_1<N_2$ and the second inequality is guaranteed by the optimality of $\hat{x}_{i \alpha}^C$ for \eqref{equ:lowerlevel} with $N_2$. Then, we rewrite the above inequality as
\[N_1 \log\sum_{i=1}^L\sum_{\alpha=1}^L r_{i\alpha}^C \tilde{x}_{i \alpha}^C - \sum_{i=1}^L \sum_{\alpha=1}^L \tau^{dr}_{i\alpha} \tilde{x}_{i\alpha}^C< \]
\[
N_1 \log\sum_{i=1}^L\sum_{\alpha=1}^Lr_{i\alpha}^C \hat{x}_{i \alpha}^C- \sum_{i=1}^L\sum_{\alpha=1}^L \tau^{dr}_{i\alpha} \hat{x}_{i\alpha}^C. 
\]
This contradicts the optimality of $\tilde{x}_{i \alpha}^C$ for \eqref{equ:lowerlevel} with $N_1$. Thus, only case iv is possible, meaning that both $\sum_{i=1}^L\sum_{\alpha=1}^Lr_{i\alpha}^C x_{i \alpha}^C$ and $\sum_{i=1}^L\sum_{\alpha=1}^L \tau^{dr}_{i\alpha} x_{i \alpha}^C$ increase in $N$. Notice that here our discussion ignores the cases of equality for simplicity since all of them follow the similar arguments. The proof is completed.

\subsection{Property (4)}

We can prove this by applying Lemma 5 of \citealt{antonis} to our particular ride-hailing system.

\section{Proof of Proposition \ref{pro:nohiding}}\label{app:nohiding}
As shown in \citealt{antonis} and discussed in Proposition \ref{pro:uniqueness}(c), all CV equilibria $x_{i \alpha}^C$ obtained by solving \eqref{equ:lowerlevel} also attains the maximum of the following problem
\begin{subequations} 
\begin{align}
\max_{\bm{x}^C} \quad & \sum_{i=1}^L\sum_{\alpha=1}^L r_{i\alpha}^C x_{i \alpha}^C \\
\text{s.t.}  \quad & \sum_{j=1}^L x_{ji}^C \le b_i^C, \quad \forall i, \label{equ:demand3}\\
& \sum_{j=1}^L (\sum_{k=1}^L x_{kj}^C) q_{ji} = \sum_{\alpha = 1}^L x_{i \alpha}^C, \quad \forall i,\\
&\sum_{i=1}^L\sum_{\alpha=1}^L \tau^{dr}_{i\alpha}x_{i\alpha}^C\leq m_0^C, \label{equ:activem2}\\
& x_{i \alpha}^C \ge 0, \quad \forall i, \alpha,
\end{align}
\end{subequations}
where $m_0^C$ denotes the CV active mass in the equilibrium. 

When there is no queueing CV, we have $m_0^C$ equal to the largest possible value $N$, the CV fleet size. Thus, if the platform hides demand in some regions from CVs, the active mass $m_0^C$ of the new equilibrium cannot be larger. Moreover, since both $b_i^C$ and $m_0^C$ become smaller in constraints \eqref{equ:demand3} and \eqref{equ:activem2}, the feasible region shrinks and the optimal objective function value of $\sum_{i=1}^L\sum_{\alpha=1}^L r_{i\alpha}^C x_{i \alpha}^C$ must be non-increasing.
Then, based on \eqref{equ:transfer}, since both the profit of CVs and the active mass are non-increasing, the platform profit earned by CVs is also non-increasing. The proof is completed.

\section{Proof of Lemma \ref{l:active}}

Notice that CV equilibrium $(\{x_{i\alpha}^C\},\{w_i^C\})$ is only determined by the revealed demand to CVs $\bm{b}^C$. Since $(\bm{x}^A,\bm{w}^A,\bm{x}^C,\bm{w}^C)$ is a mixed-fleet equilibrium, there exists a $\bm{b}^C$ such that $(\bm{x}^C,\bm{w}^C)$ is a solution of problem $\mathcal{CV}(\bm{b}^C)$. Naturally, the CV flow balance constraint and CV mass constraint are satisfied.

As for AVs, $(\bm{x}^A,\bm{w}^A)$ operates in the way that maximizes profit from serving demand $\{\sum_{j}x_{ji}^A\}_{i=1}^L$. By the revelation control, the queueing behavior of AVs and CVs are separated, where AVs serve demand $\{\sum_{j}x_{ji}^A\}_{i=1}^L$ and CVs serve demand $\bm{b}^C$. 


If we take a closer look at the AV queue, it makes no difference to let an AV wait in a queue or store in a parking lot. Hence, the platform only needs to optimally operate the active AVs and leave the remaining unutilized, which satisfies the AV mass constraint. In this way, the AVs on the road are always busy and do not wait, i.e., $\bm{w}^A = \bm{0}$.

Since $(\bm{x}^A,\bm{w}^A,\bm{x}^C,\bm{w}^C)$ and $(\bm{x}^A,\bm{0},\bm{x}^C,\bm{w}^C)$ have the same service rates, they make the same platform profit, meaning that $(\bm{x}^A,\bm{0},\bm{x}^C,\bm{w}^C)$ attains the highest total platform profit and the platform equilibrium is satisfied. Therefore, we can conclude that if $(\bm{x}^A,\bm{w}^A,\bm{x}^C,\bm{w}^C)$ is a mixed-fleet equilibrium, then under revelation control and storage of unutilized AVs, $(\bm{x}^A,\bm{0},\bm{x}^C,\bm{w}^C)$ satisfies the three conditions in Definition \ref{d:mixed equi} and is also a mixed-fleet equilibrium.

\section{Proof of Theorem \ref{thm:bilevel}} \label{app:bilevel}

Proof of 1). Suppose that $z =(\bm{x}^A,\bm{0},\bm{x}^C,\bm{w}^C)$ is an equilibrium of the mixed-fleet system. Then by the definition of mixed-fleet equilibrium, it achieves the highest possible profit for the platform in all situations where part of the demand is served by AVs and part of the remaining demand is offered to CVs. We must prove that it must also solve $\mathcal{OPT}$. Suppose this is not the case. Then, since the corresponding $(\bm{x}^A,\bm{b}_e^C)$ (to $z$) is a feasible point in $\mathcal{OPT}$, there must exist another feasible point $(\bm{x}^{A'},\bm{b}^{C'})$ in $\mathcal{OPT}$ that achieves a strictly higher profit than $(\bm{x}^A,\bm{b}_e^C)$. But this corresponds to a set of feasible rates for the AVs and an equilibrium for the CV system, i.e., an equilibrium for the mixed-fleet system achieving a higher profit than $(\bm{x}^A,\bm{0},\bm{x}^C,\bm{w}^C)$, which is a contradiction.

Proof of 2). Straightforward, by reversing the previous arguments. If a solution to $\mathcal{OPT}$ is not an equilibrium of the mixed-fleet system, then an equilibrium of the mixed-fleet system will correspond to a case in $\mathcal{OPT}$ that produces strictly higher profit, hence a contradiction.

\section{Proof of Proposition \ref{pro:allav}} \label{app:allav}

We prove the first part by arguing that when there are unutilized AVs, the platform can always increase its profit by using a small fraction of them to simulate the equilibrium strategy of CVs, i.e., have them operate like CVs. This has the equivalent effect of keeping the active AVs operating as before but increasing the fleet size of CVs. Thus, what remains is to prove the platform profit from CVs is non-decreasing in CV fleet size $N$, which holds given property (2) in Proposition \ref{pro:uniqueness}.

The second part is proved by contradiction. Suppose the optimal strategy uses both AVs and CVs to serve customers and there exists an AV strategy that can serve all demand. Then, because of the existence of CVs, all AVs must be busy transporting customers or repositioning. Note that the platform profit contains three parts: the payment from customers served by AVs, a fraction of the payment from customers served by CVs as commission fee, and (minus) the driving cost of busy AVs. The above optimal strategy attains the maximum AV driving cost since all AVs are constantly driving. However, since the commission rate $R<1$, the total collected customer payments under the optimal strategy must be smaller than using only AVs to fulfill the demand. This contradicts optimality.

\section{Detailed Discussion of Example 1} \label{app:example1}

In this example, the AV-first policy is simple: use the AVs first to serve the demand in the high profit region 1 and then serve the lower profit region 2 until all demand is served or all AVs are used, then reveal all the residual demand $b_i - \sum_{j}x_{ji}^A$ to CVs. Next, we will discuss the AV-first solution in Example 1 depending on the fleet size of $M$ and $N$:

\noindent\textbf{Case (a):} In the AV-first policy, AVs can serve all the demand in the high profit region 1, i.e., $M \ge b \tau$, and possibly some demand in the lower profit region 2. If there is residual demand in region 2 under the AV-first, CVs will serve region 2 without queueing in region 1 since there is no left demand in the high profit region as another option for them.

\noindent\textbf{Case (b):} In the AV-first policy, AVs cannot fully serve the demand in the high profit region 1, i.e., $M<b\tau$. CVs prefer to serve the residual demand in region 1 since it generates a larger rate of revenue than serving region 2. If the CV fleet size $N$ is small, CVs can serve the demand in region 1 without forming a queue, but as $N$ increases, a CV queue in region 1 will be formed. Further, when the waiting time in region 1 reaches a certain threshold they start serving region 2 (e.g. waiting time threshold is equal to the repositioning time $\tau$ if no demand is hidden from CVs). Eventually, after a queue starts being formed in region 2, they keep “water-filling” both queues in region 1 and 2. We further distinguish the following cases according to the mass $N$ of CVs:
\begin{itemize}
\item \textbf{Case (b1):} CVs are not enough to serve the residual demand in region 1, i.e., $N\le b\tau-M$. In this case, all AVs and CVs serve region 1 and there is no queue formed. 

\item \textbf{Case (b2):} CVs are enough to serve the demand in region 1, i.e., $N>b\tau -M$, but the queue formed by CVs in region 1 has a waiting time no larger than $\tau$, i,e., the number of CVs waiting in region 1 is less than the number of CVs busy serving customers in region 1, $N\le2(b\tau -M)$. By AV-first policy, all the demand is served in region 1 and there is no vehicle serving region 2. When the commission rate $R$ is large, the platform might prefer to let CVs serve region 1 to reduce CVs wasting in queueing and use the AVs to serve region 2.

\item \textbf{Case (b3):} CVs are enough to serve the demand in region 1 and form a large enough queue with waiting time $\tau$ to incentivize the repositioning from 1 to 2, i.e., $2(b\tau-M)<N$, but the repositioning rate is no larger than $M/2\tau$, i.e., $N<2b\tau-M$. Note that $M/2\tau$ is the service rate of AVs serving region 2 when all AVs reposition, which is the highest service rate AVs can achieve in region 2. In this case, all the demand in region 1 and partial demand in region 2 are served in the AV-first. The platform might prefer to let CVs serve region 1 when the commission rate $R$ is large due to the same reason as case b2 and the threshold of $R$ is relatively larger than that in case b2.

\item \textbf{Case (b4):} CVs are enough to serve the demand in region 1 and serve the demand in region 2 with a repositioning rate no less than $M/2\tau$, i.e., $N\ge 2b\tau-M$. In this case, the demand in region 2 can be served at a relatively high rate and it is not profitable for the platform to change to assign demand in region 1 to CVs as much as possible in order to serve more demand in region 2 by AVs.

\end{itemize}

In summary, we can derive that in cases a, b1 and b4, AV-first achieves optimal; in case b2, AV-first achieves optimal if and only if $R\le 1/2$; and in case b3, AV-first achieves optimal if and only if  $R\le \frac{M-(b\tau-N)^+}{b\tau-|b\tau-N|}$. 

Notice in case b2 and b3, when the commission rate $R$ is large, the optimal solution is achieved when CVs behave as if all the demand is revealed to them. To simplify the explanation, we call this revelation/solution `\textbf{CV-first policy}': CV strategy $\tilde{\bm{x}}^C$ is the solution of \eqref{equ:lowerlevel} with all the demand revealed (i.e., $\bm{b}^C=\bm{b}$). Then obtain the AV strategy $\tilde{\bm{x}}^A$ that optimizes the revenue earned by AVs over all $\bm{x}^A$ that satisfies $0\le \sum_{j=1}^L x_{ji}^A \le b_i-\sum_{j=1}^L\tilde{x}_{ji}^C, \forall i$. Note that this example is a special case where the optimal performance can always be achieved by either the AV-first or the CV-first policy, which is not true for general networks.

\section{Proof of Proposition \ref{pro:allav3}}\label{app:allav3}

As for case i, first note that if the optimal scheduling uses AVs only to serve customers, then the AV-first must be optimal. Thus, we only need to consider the situation where there are active CVs serving the residual demand under the optimal scheduling. Given that, based on Proposition \ref{pro:allav} and the definition of AV-first policy, all AVs must be non-idling under the optimal scheduling and the AV-first policy.

Assuming the payment from customers served by AVs is similarly divided into two parts with proportion $R$ (commission rate) and the residual $1-R$, we can rewrite the platform profit as

\[\Pi_{pl}= R \times \text{total customer payment to  AVs and CVs} \]
\[+(1-R) \times \text{total customer payment to AVs}\]
\[- \text{total AV driving cost}.\]

Note that all the demand is served under the AV-first policy in case i, hence its result reaches the maximum value of the first term. In addition, since all AVs are utilized under the optimal scheduling and the AV-first policy, both of them result in the third term reaching the maximum value $Mc$, meaning that all $M$ AVs drive all the time. It remains to show that the AV-first policy maximizes the second term. Notice that by the definition the profit the platform earns from AVs is equal to the customer payment to AVs minus the AV drving cost. In other words, the customer payment to AVs is actually equals to the summation of the profit the platform earns from AVs and the AV driving cost. Since the AV-first policy attains the maximum AV driving cost in this case and maximizes the profit the platform earns from AVs by its definition, it can be found that the AV-first policy also maximizes the customer payment to AVs, i.e., the second term. In conclusion, we have shown the AV-first policy is (at least) optimal.

Case ii is trivial. If there are enough AVs so that there exists an AV strategy that can serve all the demand, then by Proposition \ref{pro:allav} (2), the optimal policy must use AVs only, i.e., the optimal platform profit is equivalent to the maximum profit AVs can earn to the platform. Thus, the AV-first policy is must be optimal.

\section{Proof of Proposition \ref{pro:av-first loss}} \label{app:av-first loss}
To give a clear analysis and proof for Proposition \ref{pro:av-first loss}, we first offer an informal discussion of how to improve from the AV-first scheduling in two-region networks, then provide rigorous proof.

Consider the general two-region networks with arbitrary arrival rates where wlg $b_{12}\leq b_{21}$. There are two (out of the four possible) important policy cycles that we need to consider. Cycle 1 has the highest reward rate (continuously serve customers), with drivers always staying at the region where they deposited the last customer to pick up the next. Since it is the most preferred cycle and all other cycles have strictly lower reward rates if no queues are formed in any of the regions, drivers will first fill that cycle without queuing at any of the regions until (we omit the simple analysis) the corresponding service rates of customers in cycle 1 become $b^{(1)}_{11}=b_{11}$, $b^{(1)}_{12} =b^{(1)}_{21}= b_{12}$ and $b^{(1)}_{22}=b_{22}\frac{b_{12}}{b_{21}}$. Note that since there is no more demand in region 1, the next interesting cycle is cycle 2 that always serve region 2 where there is remaining demand, but its reward rate is lower because of repositioning.  

As more drivers join cycle 1, they will prefer to wait in region 1 until eventually both cycles become equivalent in terms of revenue. Similar to Example 1, suppose AV-first provides a situation where AVs are prioritized to serve cycle 1 but not enough to fully serve it. Then, CVs first serve the residual demand in cycle 1 and further cause waiting but not enough for drivers to choose cycle 2 where there is unserved demand. Similar as in Example 1, more customers can be served by replacing in cycle 1 AVs with CVs that would otherwise be waiting and sending the AVs to serve cycle 2. The maximum increase of served customers can be achieved if all AVs in cycle 1 are replaced by waiting CVs. As the commission rate $R\to 1$, we can show such a replacement can perform at most $20\%$ better than the AV-first.

Now, we rigorously prove Proposition \ref{pro:av-first loss}. Motivated by Example 1, we start by proving the following lemma for this special case of two-region networks.

\begin{lemma}
For any two-region network with $b_{12}=b_{22}=0$, $b_{11}, b_{21} > 0$ (similar to Example 1) and the driving cost rate $c=0$, the maximum performance loss of the AV-first policy as compared to the optimum is $20\%$, which is achieved when $M=\frac{b_{11}\tau_{11}}{2}$, $N=b_{11}\tau_{11}$, $\tau_{12}=\tau_{21}$ and $R\to 1$.
\end{lemma}


Notice that in networks like Example 1, the optimum can be achieved by either the AV-first policy or the CV-first policy (see Appendix \ref{app:example1}). Hence, our objective turns to find the maximum performance difference between the AV-first policy and the CV-first policy in cases b2 and b3. Notice that in these two cases, since the profit CVs earn to the platform under the CV-first policy is larger than that under the AV-first policy, their performance difference is enlarged as the commission rate $R$ increases. Thus, the maximum performance loss of the AV-first policy is achieved when $R\to 1$.

Given $R\to 1$, we next show that in cases b2 and b3, the maximum performance difference is achieved when CVs are enough to serve the residual demand in region 1 and form a queue with waiting time $w^*$ under the AV-first policy, i.e., $N=N^*:=(b_{11}-M/\tau_{11})(\tau_{11}+w^*)$, where $w^*$ is the minimum waiting time required for incentivizing CVs to reposition from region 1 to region 2 and is given by $\frac{\tau_{11}\tau_{12}}{\tau_{21}}$.

The intuition is as follows: i) When the number of CVs is less than $N^*$, i.e., in case b2, increasing the number of CVs will only increase the number of queuing CVs and not change the profit generated by the AV-first policy, but the profit generated by the CV-first policy must be non-decreasing as the number of CVs increases. Thus, in this case, increasing the number of CVs leads to a larger performance loss of the AV-first policy as compared to the CV-first policy. ii) When the number of CVs is larger than $N^*$, i.e., in case b3, decreasing the CV fleet size to $N^*$ under the AV-first policy will decrease the profit by $p\frac{\tau_{21}}{\tau_{12}+\tau_{21}}$ for each reduced CV since these CVs used to reposition to serve region 2. However, we will show the decreased profit under the CV-first policy cannot be larger than this. This is because when $N > N^*$, under the CV-first policy, the total number of AVs and CVs is always enough to serve all the demand in region 1. Thus, as the CV fleet size decreases to $N^*$, a profit reduction is possible only when the served demand in region 2 is reduced.

A more detailed discussion of CV-first in case ii) is as follows for two different situations. First, under the CV-first policy, if the CV fleet size is not enough to form a queue in region 1, then by decreasing the CV fleet size to $N^*$, the demand in region 1 that was served by these decreased CVs is now served by the AVs which used to reposition serving region 2. Thus, the served demand in region 2 is reduced, leading to the same profit decrease as in the AV-first policy, i.e., the profit decreases by $p\frac{\tau_{21}}{\tau_{12}+\tau_{21}}$ for each reduced CV. Second, if the CV fleet size is enough to form a queue at region 1 under the CV-first policy, decreasing the CV fleet size until the queue becomes empty does not change the profit generated by the CV-first policy. Then as we keep decreasing the CV fleet size to $N^*$, the profit decreases in the same way as in the former situation. Therefore, we can conclude that when the CV fleet size is larger than $N^*$, decreasing the number of CVs leads to a larger performance loss of the AV-first policy as compared to the CV-first policy. Summarizing cases i) and ii), the maximum performance loss of the AV-first policy is achieved when CV fleet size equals $N^*$.

Given the number of CVs $N=N^*=(b_{11}\tau_{11}-M)(1+\frac{\tau_{12}}{\tau_{21}})$, we next discuss two possible cases comparing the number of AVs($M$) and the number of queueing CVs($(b_{11}\tau_{11}-M)\frac{\tau_{12}}{\tau_{21}}$).

When $M\ge (b_{11}\tau_{11}-M)\frac{\tau_{12}}{\tau_{21}}$ (or equivalently, $\frac{M}{b_{11}\tau_{11}}\ge \frac{\tau_{12}}{\tau_{12}+\tau_{21}}$), i.e., under the AV-first policy, the number of queueing CVs is less than the number of AVs. Then, compared to the AV-first policy, in the CV-first policy, $(b_{11}\tau_{11}-M)\frac{\tau_{12}}{\tau_{21}}$ of AVs are driven out from serving demand in region 1 by CVs to serve region 2. Hence, the performance loss of the AV-first policy as compared to the CV-first policy is given by
\[1-\frac{b_{11}\tau_{11}}{b_{11}\tau_{11}+(b_{11}\tau_{11}-M)\frac{\tau_{12}}{\tau_{21}}\frac{\tau_{21}}{\tau_{12}+\tau_{21}}}\]
\[=1-\frac{1}{1+\left(1-\frac{M}{b_{11}\tau_{11}}\right)\frac{\tau_{12}}{\tau_{12}+\tau_{21}}} \leq 1-\frac{1}{1+\frac{\tau_{12}\tau_{21}}{(\tau_{12}+\tau_{21})^2}}\le \frac{1}{5},\]
where the first inequality holds since $\frac{M}{b_{11}\tau_{11}} \ge \frac{\tau_{12}}{\tau_{12}+\tau_{21}}$ and the second inequality becomes equality when $\tau_{12}=\tau_{21}$.

When $M\le (b_{11}\tau_{11}-M)\frac{\tau_{12}}{\tau_{21}}$ (or equivalently, $\frac{M}{b_{11}\tau_{11}}\le \frac{\tau_{12}}{\tau_{12}+\tau_{21}}$), i.e., under the AV-first policy, the number of queueing CVs is more than the number of AVs. Then compared to the AV-first policy, in the CV-first policy, all of $M$ AVs are driven out from serving demand in region 1 by CVs to serve region 2. Hence, the performance loss of the AV-first policy as compared to the CV-first policy is given by
\[1-\frac{b_{11}\tau_{11}}{b_{11}\tau_{11}+M\frac{\tau_{21}}{\tau_{12}+\tau_{21}}} = 1-\frac{1}{1+\frac{M}{b_{11}\tau_{11}}\frac{\tau_{21}}{\tau_{12}+\tau_{21}}}\]
\[\leq 1-\frac{1}{1+\frac{\tau_{12}\tau_{21}}{(\tau_{12}+\tau_{21})^2}}\le \frac{1}{5}, \]
where the first inequality holds since $\frac{M}{b_{11}\tau_{11}} \le \frac{\tau_{12}}{\tau_{12}+\tau_{21}}$ and the second inequality becomes equality when $\tau_{12}=\tau_{21}$.

In summary, the maximum performance loss of the AV-first is achieved when $R\to 1$, $\tau_{12}=\tau_{21}$, $M =b_{11}\tau_{11} \frac{\tau_{12}}{\tau_{12}+\tau_{21}}=\frac{b_{11}\tau_{11}}{2}$ and $N =(b_{11}\tau_{11}-M)(1+\frac{\tau_{12}}{\tau_{21}})= b_{11}\tau_{11}$.

Now coming back to Proposition \ref{pro:av-first loss}, let us consider a general two-region network with four possible routes ($1\to 1$, $1\to 2$, $2\to 1$ and $2\to 2$,). Without loss of generality, we assume there is imbalanced cross-region demand, $b_{12} \le b_{21}$. 

Similar to the two-routes network we just analyzed, for any two-region network with imbalanced demand between the two regions, the customer demand can be decomposed into two service cycles for vehicles to circulate: demand $b^{nor}$ that can be served by vehicles with the deterministic policy that always choose to serve the local customers when they become empty, and demand $b^{rep}$ that can be served by vehicles with the deterministic policy that choose to reposition when becoming empty in region 1 with the lower demand and serve local customer when becoming empty in region 2:
\[b^{nor}=\left[\begin{array}{cc}
   b_{11}& b_{12}\\
   b_{12} & b_{12}\frac{q_{22}}{q_{21}}
    \end{array}
    \right],\]
\[b^{rep}=\left[\begin{array}{cc}
   0 & 0\\
   b_{21}-b_{12} & (b_{21}-b_{12}) \frac{q_{22}}{q_{21}}
    \end{array}
    \right].\]
Notice that, in the first cycle with demand $b^{nor}$, vehicles are busy serving customers all the time and have the maximum reward rate, while vehicles in the second cycle have a smaller reward rate due to the additional repositioning driving/time cost. Hence, similar to what we have discussed for Example 1, self-interested CVs will serve the `lower profit' second circle only if the average reward from the first circle reduces to be equal to the reward of the cycle of serving always region 2 due to the formed queue in region 1. Then, by using similar analysis, we can prove the maximum performance loss of the AV-first policy is $20\%$ the reward rate of the second cycle is half as compared to the first cycle, the fleet size of AVs is half of that required to serve all the demand in the first cycle and the number of CVs is equal to the required number such that the remaining demand in the first cycle can be fully served but none of the demand in the second cycle being served.

\section{Analysis for General Two-region Networks}\label{app:tworegion}

In ride-hailing studies, a two-region network (i.e., $L=2$) is the simplest case to capture the spatial variability of supply and demand and provide insights on how vehicles deal with the imbalance in equilibrium \citealt{purecv,av}. As in \citealt{av}, we consider a two-region, four-route network (two within-region routes and two cross-region routes). We assume there are imbalanced cross-region demand, $b_{12} \neq b_{21}$, and without loss of generality, $b_{21}>b_{12}\geq 0$. In this appendix, by developing a new network splitting method, we analyze how AV fleet size $M$ and the CV fleet size $N$ influence the equilibrium of the mixed-fleet problem. Finally, we derive the closed-form CV strategy $\bm{x}^{C\star}$ and AV strategy $\bm{x}^{A\star}$ induced by the optimal solution $\bm{b}^{C\star}$ to $\mathcal{OPT}$.

\subsection{Network Splitting}\label{sec:split}
Similar to the specific case of Example 1, for any two-region network with imbalanced demand between the two regions, the customer demand can be decomposed into two service cycles for vehicles to circulate: one served by vehicles with the deterministic policy that always chooses to serve the local customers when they become empty and the another served by vehicles with the deterministic policy that chooses to reposition to region 2 with higher demand when becoming empty in region 1 and serve local customer when becoming empty in region 2. In the first cycle, vehicles are busy serving customers all the time and have the maximum reward rate, while vehicles in the second cycle have a smaller reward rate due to the additional repositioning driving/time cost. Then, based on the balance constraints, we can infer the normalized service rates in the two cycles as follows:
\[x^{nor}=\left[\begin{array}{cc}
   1& 0\\
   0 & \frac{q_{12}}{q_{21}}
    \end{array}
    \right],\]
\[x^{rep}=\left[\begin{array}{cc}
   0 & 1\\
   0 & \frac{q_{22}}{q_{21}}
    \end{array}
    \right].\]
Notice that the service rates of AVs and CVs can always be expressed as a linear combination of $x^{nor}$ and $x^{rep}$. Let $m^{nor}=q_{11}\tau_{11}+q_{12}\tau_{12}+q_{12}\tau_{21}+\frac{q_{12}q_{22}}{q_{21}}\tau_{22}$ denote the mass of vehicles required to serve customers at rate $x^{nor}$ and $m^{rep}=\tau_{12}+\tau_{21}+\frac{q_{22}}{q_{21}}\tau_{22}$ denote the mass of vehicles required to serve customers at rate $x^{rep}$. Moreover, we define $r^{nor}_A$ ($r^{nor}_{C2P}$) and $r^{rep}_A$ ($r^{rep}_{C2P}$) to denote the average platform profit earned by per unit mass of AVs (CVs) that use the strategy associated with rate $x^{nor}$ and $x^{rep}$ (in other words, serve the first cycle and the second cycle), respectively:
\begin{subequations}
\begin{align}
r^{nor}_A=&p-c,\nonumber\\
r^{rep}_A=&\gamma_{1} p-c,\nonumber\\
r^{nor}_{C2P}=&Rp,\nonumber\\
r^{rep}_{C2P}=&\gamma_{1} Rp,\nonumber\\
r^{nor}_{C}=&r^{nor}_A-r^{nor}_{C2P},\nonumber\\
r^{rep}_{C}=&r^{rep}_A-r^{rep}_{C2P},\nonumber
\end{align}
\end{subequations}
where $\gamma_{1}:=\frac{\tau_{21}+\tau_{22}\frac{q_{22}}{q_{12}}}{\tau_{12}+\tau_{21}+\tau_{22}\frac{q_{22}}{q_{21}}}<1$ denotes the fraction of the time of serving customer in the total active time (serving customers or repositioning) when serving the second cycle. Since $\gamma_{1}<1$, it is easy for us to find, for both AVs and CVs, it is more profitable to serve the first cycle than serve the second cycle. Moreover, for each cycle, whether or not it is profitable for CVs to serve depends on the network parameters and commission rate $R$. More specifically, we can obtain the following thresholds for the commission rate $R$:
\begin{subequations}\label{equ:thresholds}
\begin{align}
R_{th1}=&1-\frac{c}{p},\\
R_{th2}=&1-\frac{c}{\gamma_{1}p}.
\end{align}
\end{subequations}
When $0<R\leq R_{th2}$, we have $r^{nor}_A\geq r^{nor}_{C2P}$ and $r^{rep}_A\geq r^{rep}_{C2P}$, and thus $r^{nor}_{C}\geq 0$ and $r^{rep}_{C}\geq 0$, i.e., it is profitable for CVs to serve both cycles. As $R$ increases to be larger than $R_{th2}$, we have $r^{nor}_A\geq r^{nor}_{C2P}$ and $r^{rep}_C<0$, i.e., the commission fee received by CVs from serving the second cycle cannot cover the total driving cost of serving customers and repositioning from region 1 to region 2. Further, as $R$ keeps increasing to be larger than $R_{th1}$, we have $r^{nor}_C<0$ and $r^{rep}_C<0$ and thus CVs will never serve any demand. In this appendix, we only discuss the most complicated case of $R\leq R_{th2}$ where CVs might serve both cycles. Other cases are simpler and can use the similar analysis.

\subsection{Closed-Form Solutions}

Next, we discuss how AV fleet size $M$ and CV fleet size $N$ jointly determine AV and CV strategies depending on the thresholds given in \eqref{equ:thresholds}.

First let $m_{th1}:=(b_{11}+b_{12})m^{nor}$ denote the number of vehicles that is needed for serving all the demand $b_{11}+b_{12}$ in region 1 using the strategy associated with the $x^{nor}$. Let $m_{th2}:=(b_{21}-b_{12} )m^{rep}$ denote the number of vehicles that is needed for serving all the imbalanced demand $b_{21}-b_{12}$ using the strategy associated with the $x^{rep}$. Let $w^*$ denote the threshold above which the queue (in terms of waiting time) is long enough to incentivize the CVs waiting in region $1$ to reposition to region $2$, which should satisfy:
\[\frac{m_{th1}((1-R)p-c)}{(b_{11}+b_{12})w^*+m_{th1}}=\gamma_{1}(1-R)p-c,\]
and thus $w^*$ is given by
\[w^*=\frac{\frac{m_{th1}((1-R)p-c)}{\gamma_{1}(1-R)p-c}-m_{th1}}{b_{11}+b_{12}}.\]

Then we can use $m_{th3}:=m_{th1}+w^*(b_{11}+b_{12})=\frac{((1-R)p-c)}{\gamma_{1}(1-R)p-c}m_{th1}$ to denote the minimum number of CVs that is needed for starting to serve the second cycle with extra repositioning from region 1 to region $2$ given all the demand is revealed to CVs. Furthermore, we denote $\gamma_{2}:=\frac{\gamma_{1}(1-R)p-c}{((1-R)p-c)}<\gamma_{1}$ and thus we have $\gamma_{2}m_{th3}=m_{th1}$

Then, we can discuss the optimality of the AV-first policy in the following cases:

\noindent \textbf{Case (a):} There are enough AVs to serve all the demand in region 1 using the strategy associated with $x^{nor}$, i.e., $M\geq m_{th1}$. In this case, when using the AV-first policy, AVs serve the first cycle with the highest rate $(b_{21}-b_{12})x^{nor}$ and CVs will serve the second cycle without forming a queue in region 1 since there is no left demand in region 1. Thus, the AV-first policy is optimal and we have
\begin{subequations}
\begin{align}
\bm{x}^{A\star}=& (b_{11}+b_{12})x^{nor}\nonumber + \min((b_{21}-b_{12}),\frac{M-m_{th1}}{m^{rep}})x^{rep}, \nonumber
\end{align}
\end{subequations}
\[\bm{x}^{C\star}=\min(\frac{N}{m^{rep}},((b_{21}-b_{12})-\frac{M-m_{th1}}{m^{rep}})^+) x^{rep},\]
and the optimal total platform profit is given by
\[m_{th1}r^{nor}_{A} +(M-m_{th1}) r^{rep}_{A}+\min(N,(m_{th1}+m_{th2}-M)^+)r^{rep}_{C}.\]

\noindent \textbf{Case (b):} AVs and CVs, together, cannot fully serve the demand in region 1 using the strategy associated with $x^{nor}$, i.e., $N+M\le m_{th1}$. In this case, it is optimal to use all AVs and CVs to serve the first cycle as in the AV-first policy and we have
\[\bm{x}^{A\star}=\frac{M}{m^{nor}}x^{nor},\]
\[\bm{x}^{C\star}=\frac{N}{m^{nor}}x^{nor},\]
and the optimal total platform profit is given by
\[Mr^{nor}_{A} +N r^{nor}_{C}.\]

\noindent \textbf{Case (c):} AVs cannot fully serve fully serve the demand in region 1 using the strategy associated with $x^{nor}$ (i.e., $M<m_{th1}$), but the number of CVs are enough serve all the residual demand in region 1 and region 2 (i.e., $N\ge (m_{th1}-M)/\gamma_{2}+m_{th2}$). Notice that based on Proposition \ref{pro:allav3}, the AV-first policy must be optimal if the residual demand can be fully served by CVs in the CV equilibrium. Thus, in this case, the AV-first policy is optimal and we have
\[\bm{x}^{A\star}=\frac{M}{m^{nor}}x^{nor},\]
\begin{subequations}
\begin{align}
\bm{x}^{C\star}=\frac{m_{th1}-M}{m^{nor}}x^{nor} + (b_{21}-b_{12})x^{rep}, \nonumber
\end{align}
\end{subequations}
and the optimal total platform profit is given by
\[Mr^{nor}_{A} + (m_{th1}-M)r^{nor}_{C}+ m_{th2} r^{rep}_{C}.\]

\noindent \textbf{Case (d):} the remaining case is with $M<m_{th1}$ and $m_{th1}-M<N<(m_{th1}-M)/\gamma_{2}+m_{th2}$. In this case, depending on whether or not the platform will choose to hide demand from CVs in the optimal solution, we can discusses the following two situations:

1) In the optimal solution, the platform needs to hide demand from CVs to induce the better equilibrium. In this case, we first notice that, as discussed in Section \ref{sec:cvproperty}, the intuition behind the platform hiding demand from CVs is that the platform can incentivize CVs that queue in a given region to reposition by reducing the demand in that region and increasing the waiting time. In our considered two-region network, the only possible vehicle reposition is from region 1 to region 2. Thus, the only possible region the platform choose to hide demand is region 1 to induce CVs to serve less the first cycle and more the second cycle with the repositioning. Then, given the platform hide demand in region 1 in the optimal solution, all the AVs must be used to serve the first cycle otherwise the platform can increases the profit by assigning AVs who serve the second cycle to serve the hidden part of the first cycle. Thus, we have
\begin{subequations}
\begin{align}
\bm{x}^{A}_{hide}=& \frac{M}{m^{nor}}x^{nor}\nonumber
\end{align}
\end{subequations}
and the residual demand that is unserved by AVs is
\[b_{res}=\left[\begin{array}{cc}
   (1-\frac{M}{m_{th1}})b_{11}& (1-\frac{M}{m_{th1}})b_{12}\\
   b_{21}-\frac{M}{m_{th1}}b_{12} & b_{22}-\frac{M}{m_{th1}}b_{12}\frac{b_{22}}{b_{21}}
    \end{array}
    \right],\]
Notice that CVs will serve demand in region 2 with higher priority. Let $m_{th4}:=(b_{21}-\frac{M}{m_{th1}}b_{12})m^{rep}$ denote the number of vehicles required to serve all the demand in region 2 using the strategy of the second cycle. Then, we have
\begin{subequations}
\begin{align}
\bm{x}^{C}_{hide}=& \frac{\min(m_{th4},N)}{m^{rep}} x^{rep}\nonumber\\ 
  &+ \frac{(N-m_{th4})^+}{b_{11}t_{11}+m_{th1}/\gamma_{2}-m_{th1}}\left[\begin{array}{cc}
   b_{11}+b_{12}& -b_{12}\\
   0 & b_{12}
    \end{array}
    \right]\nonumber
\end{align}
\end{subequations}
Then, the incurred optimal platform profit, denoted as $R_{hide}$, is given by
\[R_{hide}=Mr^{nor}_{A}+(N-\frac{m_{th1}(N-m_{th4})^+}{(b_{11}t_{11}+m_{th1}/\gamma_{2}-m_{th1})\gamma_{2}})r^{rep}_{C2P}\]
\[+\frac{(N-m_{th4})^+}{(b_{11}t_{11}+m_{th1}/\gamma_{2}-m_{th1})}m_{th1}r^{nor}_{C2P}.\]

2) In the optimal solution, the platform does not need to hide demand from CVs to induce the better equilibrium. We first notice that given the platform does not hide demand, the AV-first policy must be optimal $M<m_{th1}$ and $(m_{th1}-M)/\gamma_{2}+\min(m_{th2},M)\le N<(m_{th1}-M)/\gamma_{2}+m_{th2}$ since there exists no other method that can serve more demand than the AV-first. The corresponding optimal solution obtained by the AV-first policy is 
\[\bm{x}^{A}_{AV-F}=\frac{M}{m^{nor}}x^{nor},\]
and
\begin{subequations}
\begin{align}
\bm{x}^{C}_{AV-F}=& ((b_{11}+b_{12})-\frac{M}{m^{nor}})x^{nor}\nonumber\\ 
  & + (\frac{N-(m_{th1}-M)/\gamma_{2}}{m^{rep}})^+x^{rep}, \nonumber
\end{align}
\end{subequations}
Then, the incurred optimal platform profit, denoted as $R_{AV-F}$, is given by
\[R_{AV-F}=Mr^{nor}_{A}+(m_{th1}-M)r^{nor}_{C2P}\]
\[+(N-(m_{th1}-M)/\gamma_{2})^+r^{rep}_{C2P}.\]

In the remaining case with $M<m_{th1}$ and $m_{th1}-M< N<(m_{th1}-M)/\gamma_{2}+min(m_{th2},M)$, the optimal solution is achieved by either the AV-first policy or the method that first assigns CVs to serve the first cycle with priority until all demand in this cycle is served, all CVs are used or the left demand (demand in the residual first cycle and the second cycle) are not enough for AVs to serve, then assign AVs to serve the left demand. If it is the case that the AV-first policy is optimal, then the optimal solution is and the corresponding optimal platform profit is as above.

If it is the case that the other method is optimal, the optimal solution is given by
\begin{subequations}
\begin{align}
\bm{x}^{A}_{CV-F}=& \max(\frac{(M-m_{th2})^+}{m^{nor}},(b_{11}+b_{12})-\frac{N}{m^{nor}})  x^{nor}\nonumber\\ 
  & + \frac{\min(\min(M,m_{th2}),M+N-m_{th1})}{m^{rep}}x^{rep}, \nonumber
\end{align}
\end{subequations}
and
\[\bm{x}^{C}_{CV-F}= \min((b_{11}+b_{12})-\frac{(M-m_{th2})^+}{m^{nor}},\frac{N}{m^{nor}})x^{nor}.\]
Then, the incurred optimal platform profit, denoted as $R_{CV-F}$, is given by
\[R_{CV-F}=\max((M-m_{th2})^+,m_{th1}-N)r^{nor}_{A}+\]
\[\min(\min(M,m_{th2}),M+N-m_{th1})r^{rep}_A\]
\[+\min(m_{th1}-(M-m_{th2})^+,N)r^{nor}_{C2P}.\]

Finally, by comparing $R_{hide}$, $R_{CV-F}$, and $R_{AV-F}$, we can determine the corresponding optimal solution in this case.

\subsection{driving cost rate $c=0$}
When the driving cost rate $c=0$, we can find $\gamma_{1}=\gamma_{2}$ and there is no incentive for the platform to hide demands from CVs. Thus, in case (d) with with $M<m_{th1}$ and $m_{th1}-M<N<(m_{th1}-M)/\gamma_{2}+m_{th2}$, we only need to consider comparing $R_{CV-F}$, and $R_{AV-F}$ to determine which policy is optimal. More specifically, we have

\textbf{Case (d1):} CVs are enough to serve the residual demand that is not served by AVs in the first cycle, i.e., $N>m_{th1}-M$, but the queue formed by CVs in region 1 has a waiting time no larger than $w^*$, i,e., $N\le (m_{th1}-M)/\gamma_{2}$. In this case, the the queue formed by CVs is not long enough to incentivize the repositioning and thus the second cycle cannot be served when using the AV-first policy. By comparing $R_{CV-F}$, and $R_{AV-F}$, we show that the AV-first policy is optimal when $R\le 1-\gamma_{1}$ and we have

\[\bm{x}^{A\star}=\frac{M}{m^{nor}}x^{nor},\]
and 
\[\bm{x}^{C\star}=((b_{11}+b_{12})-\frac{M}{m^{nor}})x^{nor}.\]

The CV-first policy is optimal when $R\ge 1-\gamma_{1}$ and we have
\begin{subequations}
\begin{align}
\bm{x}^{A\star}=& \max(\frac{(M-m_{th2})^+}{m^{nor}},(b_{11}+b_{12})-\frac{N}{m^{nor}})  x^{nor}\nonumber\\ 
  & + \frac{\min(\min(M,m_{th2}),M+N-m_{th1})}{m^{rep}}x^{rep}, \nonumber
\end{align}
\end{subequations}
and
\[\bm{x}^{C\star}= \min((b_{11}+b_{12})-\frac{(M-m_{th2})^+}{m^{nor}},\frac{N}{m^{nor}})x^{nor}.\]

\textbf{Case (d2):} CVs are enough to serve the demand in region 1 and form a large enough queue with waiting time $w^{\star}$ to incentivize the repositioning from 1 to 2, i.e., $N>(m_{th1}-M)/\gamma_{2}$, but the number of residual CVs is no larger than $M$ and $m_{th2}$, i.e., $N\le(m_{th1}-M)/\gamma_{2}+\min(M,m_{th2})$. In this case, the AV-first policy is optimal when $R\le \frac{M-\max(m_{th1}-N,(M-m_{th2})^+)-\gamma_{1}\min(\min(M,m_{th2}),M+N-m_{th1})}{m_{th1}-\gamma_1 N-\max(m_{th1}-N,(M-m_{th2})^+)}$ and we have
\[\bm{x}^{A\star}=\frac{M}{m^{nor}}x^{nor},\]
and
\begin{subequations}
\begin{align}
\bm{x}^{C\star}=& ((b_{11}+b_{12})-\frac{M}{m^{nor}})x^{nor}\nonumber\\ 
  & + \frac{N-(m_{th1}-M)/\gamma_{2}}{m^{rep}}x^{rep}. \nonumber
\end{align}
\end{subequations}

The CV-first policy is optimal when $R\ge \frac{M-\max(m_{th1}-N,(M-m_{th2})^+)-\gamma_{1}\min(\min(M,m_{th2}),M+N-m_{th1})}{m_{th1}-\gamma_1 N-\max(m_{th1}-N,(M-m_{th2})^+)}$ and we have
\begin{subequations}
\begin{align}
\bm{x}^{A\star}=& \max(\frac{(M-m_{th2})^+}{m^{nor}},(b_{11}+b_{12})-\frac{N}{m^{nor}})  x^{nor}\nonumber\\ 
  & + \frac{\min(\min(M,m_{th2}),M+N-m_{th1})}{m^{rep}}x^{rep}, \nonumber
\end{align}
\end{subequations}
and

\[\bm{x}^{C\star}= \min((b_{11}+b_{12})-\frac{(M-m_{th2})^+}{m^{nor}},\frac{N}{m^{nor}})x^{nor}.\]

\textbf{Case (d3):} CVs are enough to serve the demand in region 1 and serve the demand in region 2 with total mass no less than $\min(M,m_{th2})$, i.e., $N\ge(m_{th1}-M)/\gamma_{2}+\min(M,m_{th2})$. In this case, the AV-first policy is optimal and we have

\[\bm{x}^{A\star}=\frac{M}{m^{nor}}x^{nor},\]
and
\begin{subequations}
\begin{align}
\bm{x}^{C\star}=& ((b_{11}+b_{12})-\frac{M}{m^{nor}})x^{nor}\nonumber\\ 
  & + \frac{N-(m_{th1}-M)/\gamma_{2}}{m^{rep}}x^{rep}, \nonumber
\end{align}
\end{subequations}

\section{Proof for Endogenous Case}\label{app:endo}

\subsection{Network Splitting}
In the imbalanced $L=2$ network, we can divide the demand into two types: i) demand $b^{nor}$ that can be served without any repositioning, and ii) demand $b^{rep}$ that needs repositioning to serve, given by 
\[b^{nor}=\left[\begin{array}{cc}
   b_{11}& b_{12}\\
   b_{12} & b_{12}\frac{q_{22}}{q_{21}}
    \end{array}
    \right],\]
\[b^{rep}=\left[\begin{array}{cc}
   0 & 0\\
   b_{21}-b_{12} & (b_{21}-b_{12}) \frac{q_{22}}{q_{21}}
    \end{array}
    \right].\]
Note that it is more profitable to serve demand $b^{nor}$ than serve demand $b^{rep}$, especially for AVs that need the platform to pay additional driving and purchase costs for repositioning mass of AVs, while the platform does not need to afford CV's repositioning/driving cost. For each type of demand, whether or not it is more profitable to serve it with AVs than with CVs depends on the AV purchase cost. To show that, we use $r^{nor}_A$ ($r^{nor}_C$) and $r^{rep}_A$ ($r^{nor}_C$) to denote the average profit earned by per unit mass of AVs (CVs) from serving demand $b^{nor}$ and $b^{rep}$, respectively:
\begin{subequations}
\begin{align}
r^{nor}_A=&p-c_A-I,\nonumber\\
r^{rep}_A=&\frac{\tau_{21}+\tau_{22}\frac{q_{22}}{q_{12}}}{\tau_{21}+\tau_{21}+\tau_{22}\frac{q_{22}}{q_{21}}}p-c_A-I,\nonumber\\
r^{nor}_C=&Rp,\nonumber\\
r^{rep}_A=&\frac{\tau_{21}+\tau_{22}\frac{q_{22}}{q_{12}}}{\tau_{21}+\tau_{21}+\tau_{22}\frac{q_{22}}{q_{21}}}Rp,\nonumber
\end{align}
\end{subequations}
By comparing the above profits, we can obtain the following thresholds for the AV purchase cost $I$:
\begin{subequations}\label{equ:thresholds2}
\begin{align}
I_{th1}=&\frac{\tau_{21}+\tau_{22}\frac{q_{22}}{q_{21}}}{\tau_{21}+\tau_{22}\frac{q_{22}}{q_{21}}+\tau_{12}}(1-R)p-c_A,\\
I_{th2}=&\frac{\tau_{21}+\tau_{22}\frac{q_{22}}{q_{21}}}{\tau_{21}+\tau_{22}\frac{q_{22}}{q_{21}}+\tau_{12}}p-c_A,\\
I_{th3}=&(1-R)p-c_A,\\
I_{th4}=&(1-R)p-c_A+\frac{\tau_{21}+\frac{q_{22}}{q_{21}}\tau_{22}}{\tau_{21}+\tau_{12}+\frac{q_{22}}{q_{21}}\tau_{22}}Rp.
\end{align}
\end{subequations}
We can observe that when the AV purchase cost $I\leq I_{th1}$, we have $r^{nor}_A\geq r^{nor}_C$ and $r^{rep}_A\geq r^{rep}_C$, and thus it is optimal for the platform to serve both demand $b^{nor}$ and $b^{rep}$ with AVs. As $I$ increases to be larger than $I_{th1}$, we have $r^{rep}_A< r^{rep}_C$ and thus it becomes better for the platform to serve the demand $d^{rep}$ that requires vehicle repositioning by CVs instead of AVs. Further, as $I$ keeps increasing to be larger than $I_{th2}$, we have $r^{rep}_A<0$ and thus the profit becomes negative for the platform to serve the demand $b^{rep}$ by AVs. Moreover, when $I>I_{th3}$\footnote{The relationship between these thresholds depends on the parameters of the model and we only have $I_{th4}>I_{th3},I_{th2}>I_{th1}$ for sure. Hereafter, we consider the case of $I_{th3}\ge I_{th2}$ in the main text, and the case of $I_{th3}<I_{th2}$ can be discussed similarly.}, we have $r^{nor}_A< r^{nor}_C$ and $r^{rep}_A< 0<r^{rep}_C$, it becomes better for the platform to serve both demand $b^{rep}$ and $b^{nor}$ by CVs instead of AVs. Finally, as $I$ keeps increasing, there exists the highest threshold $I_{th4}$ for $I$, above which it becomes better for the platform to serve the demand $b^{nor}$ by CVs instead of AVs even when the demand $b^{rep}$ is wasted.

Note that we assume the maximum possible value of AV purchase cost is $I_{max}=p-c_A$, otherwise, the platform does not procure any AVs. All the above thresholds are smaller than $I_{max}$.

\subsection{Closed-Form Solutions}

Next, we discuss how AV purchase cost $I$ and CV fleet size $N$ jointly determine AV and CV strategies depending on the thresholds given in \eqref{equ:thresholds2}.

\subsubsection{Case I}
When $I<I_{th1}$, as we have discussed earlier, it is optimal for the platform to serve all the demand by AVs, i.e., 

\[\bm{x}^{A\star}=\left[\begin{array}{cc}
   b_{11}+b_{12}& b_{21}-b_{12}\\
   0 &b_{12}+b_{22}
    \end{array}
    \right],\]
\[\bm{x}^{C\star}=\textbf{0}.\]

\subsubsection{Case II}
When $I_{th1}<I<I_{th2}$, CVs start to serve the demand $b^{rep}$ as many as possible, and each of them obtains a fixed profit $r_1:=\frac{(1-R)p(\tau_{21}+\tau_{22}\frac{q_{22}}{q_{21}})}{\tau_{12}+\tau_{21}+\tau_{22}\frac{q_{22}}{q_{21}}}-c_C$. Let $N^P(r)=\frac{rN_{max}}{((1-R)p-c_C)}$ denote the number of participating CVs if the expected CV profit is $r$. Then, the number of participating CVs is given by $N^P(r_1)$ in this case.

Note that we need $m_{th1}:=(b_{21}-b_{12} )(\tau_{12}+\tau_{21}+\tau_{22}\frac{q_{22}}{q_{21}})$ CVs $N$ to serve all the demand $b^{rep}$, and thus we have
\[\bm{x}^{C\star}=\min(1,\frac{N^P(r_1)}{m_{th1}})\left[\begin{array}{cc}
   0 & b_{21}-b_{12} \\
   0 & b_{22}-b_{12}\frac{q_{22}}{q_{21}}
    \end{array}
    \right],\]
and \[\bm{x}^{A\star}=\left[\begin{array}{cc}
   b_{11}+b_{12}& b_{21}-b_{12}\\
   0 &b_{12}+b_{22}
    \end{array}
    \right]-\bm{x}^C.\]

\subsubsection{Case III}
When $I_{th2}<I<I_{th3}$, CVs serve the demand $b^{rep}$ as many as possible and AVs only serve the demand $b^{nor}$, and each of them obtains a fixed profit $r_1$ as in case II. Then, the number of participating CVs is given by $N^P(r_1)$. Thus, we have
\[\bm{x}^{C\star}=\min(1,\frac{N^P(r_1)}{m_{th1}})\left[\begin{array}{cc}
   0 & b_{21}-b_{12} \\
   0 & b_{22}-b_{12}\frac{q_{22}}{q_{21}}
    \end{array}
    \right],\]
and \[\bm{x}^{A\star}=\left[\begin{array}{cc}
   b_{11}+b_{12}& 0\\
   0 &b_{12}+b_{12}\frac{q_{22}}{q_{21}}
    \end{array}
    \right].\]

\subsubsection{Case IV}
When $I_{th3}<I<I_{th4}$, CVs serve both the demand $b^{rep}$ and $b^{nor}$ as many as possible, and AVs serve the residual demand $b^{nor}$. But, note that only when the number of CVs is enough to serve $b^{rep}$ (i.e., $N^P\leq m_{th1}$), CVs start to serve the demand $b^{nor}$ as many as possible. Thus, each of the CVs also obtains the fixed profit $r_1$ as in cases I and II, and the number of participating CVs is $N^P(r_1)$.

\noindent\emph{Case IV-A:} if the number of CVs $N^P(r_1)$ is too small to serve all the demand $b^{rep}$ (i.e., $N^P(r_1)<m_{th1})$), we have the same $\bm{x}^{A\star}$ and $\bm{x}^{C\star}$ as in case III. 

\noindent\emph{Case IV-B:} if the number of CVs $N$ is enough to serve all the demand $b^{rep}$ (i.e., $N^P(r_1)\geq m_{th1}$), then the additional CVs will be assigned to serve the demand $b^{nor}$. Note that we need  $m_{th2}:=b_{11}\tau_{11}+b_{12}\tau_{12}+b_{21}\tau_{21}+b_{22}\tau_{22}+(b_{21}-b_{12})\tau_{12}+(b_{11}+b_{12})w^*$ to serve all the demand $b^{rep}$ and $b^{nor}$, and thus we have
\begin{subequations}
\begin{align}
\bm{x}^{C\star}\hspace{-3pt}=\hspace{-3pt}& \left[\begin{array}{cc}
   0 & b_{21}-b_{12} \\
   0 & b_{22}-b_{12}\frac{q_{22}}{q_{21}}
    \end{array}
    \right]\nonumber\\ 
  &  \hspace{-3pt}+\hspace{-3pt}\max(0,\hspace{-2pt}\min(1,\hspace{-2pt}\frac{N^P\hspace{-2pt}(r_1\hspace{-2pt})\hspace{-3pt}-\hspace{-3pt}m_{th1}}{m_{th2}\hspace{-3pt}-\hspace{-3pt}m_{th1}}\hspace{-3pt})\hspace{-1pt})\hspace{-3pt} \left[\begin{array}{cc}
   b_{11}\hspace{-3pt}+\hspace{-3pt}b_{12}& 0\\
   0 &b_{12}\hspace{-3pt}+\hspace{-3pt}b_{12}\frac{q_{22}}{q_{21}}
    \end{array}
    \hspace{-3pt}\right]\hspace{-3pt}, \nonumber
\end{align}
\end{subequations}
and \[\bm{x}^{A\star}\hspace{-3pt}=\hspace{-3pt} \min(1,\hspace{-2pt}\max(0,\hspace{-1pt}\frac{m_{th2}\hspace{-3pt}-\hspace{-3pt}N^P\hspace{-2pt}(r_1\hspace{-2pt})}{m_{th2}\hspace{-3pt}-\hspace{-3pt}m_{th1}})\hspace{-1pt})\hspace{-3pt} \left[\begin{array}{cc}
   b_{11}\hspace{-3pt}+\hspace{-3pt}b_{12}& 0\\
   0 &b_{12}\hspace{-3pt}+\hspace{-3pt}b_{12}\frac{q_{22}}{q_{21}}
    \end{array}
    \hspace{-3pt}\right]\hspace{-3pt}.\]

\subsubsection{Case V}
When $I_{th4}<I$, CVs serve both the demand $b^{rep}$ and $b^{nor}$ as many as possible. In this case, we should note that it becomes better for the platform to serve the demand $b^{nor}$ by CVs instead of AVs even when the demand $b^{rep}$ is wasted, and we need $m_{th3}:=b_{11}\tau_{11}+b_{12}\tau_{12}+b_{12}\tau_{21}+b_{12}\frac{q_{22}}{q_{21}}\tau_{22}$ of vehicles to serve all the demand $b^{nor}$. But, if the number of CVs is much larger than $m_{th3}$ to incentivize the repositioning (i.e., $N^P(r_1)\geq m_{th4}:=\max(m_{th3}\frac{I+c_A-(1-R)p}{Rp}\frac{\tau_{21}+\tau_{12}+\frac{q_{22}}{q_{21}}\tau_{22}}{\tau_{21}+\frac{q_{22}}{q_{21}}\tau_{22}},m_{th2}-\frac{(m_{th2}-m_{th1})Rp((b_{21}-b_{12})(\tau_{21}+\frac{q_{22}}{q_{21}}\tau_{22}))}{m_{th3}(I+c_A-(1-R)p)})$), CVs will choose to serve the demand $b^{rep}$ as many as possible and then serve $b^{nor}$. 

\noindent\emph{Case V-A:} if the number of CVs $N$ is too small to incentivize the repositioning (i.e., $N^P(r_1)<m_{th4}$), then CV individual profit is given by $r_2:=(1-R)*p-c_C$ as there are no waiting CVs. If $N^P(r_2)\ge m_{th3}$, all the demand $b^{nor}$ will be served by participating CVs and if $N^P(r_2)<m_{th3}$, the number of participating CVs is $N^P(r_2)$ and they are not enough to serve $b^{nor}$. AVs will serve the residual $b^{nor}$. Thus, we have 
\[\bm{x}^{C\star}=\min(1,\frac{N^P(r_2)}{m_{th3}} )\left[\begin{array}{cc}
   b_{11}+b_{12}& 0\\
   0 &b_{12}+b_{12}\frac{q_{22}}{q_{21}}
    \end{array}
    \right],\]
and \[\bm{x}^{A\star}=\max(0,1-\frac{N^P(r_2)}{m_{th3}} )\left[\begin{array}{cc}
   b_{11}+b_{12}& 0\\
   0 &b_{12}+b_{12}\frac{q_{22}}{q_{21}}
    \end{array}
    \right].\]

\noindent\emph{Case V-B:} if the number of CVs $N$ is large enough incentivize the repositioning (i.e., $N^P(r_1)\geq m_{th4}$), it is the same as case IV.

As for the case $I_{th4}>I_{th2}>I_{th3}>I_{th1}$, we define new case I with $I<I_{th1}$, case II with $I_{th1}<I<I_{th3}$, case III with $I_{th3}<I<I_{th2}$, case IV with $I_{th2}<I<I_{th4}$ and case V with $I>I_{th4}$. Note that the new cases I, II, IV, and V are the same as the above old ones, and the new case III with $I_{th3}<I<I_{th2}$ needs to be further discussed.

\subsubsection{New Case III}
When $I_{th3}<I<I_{th2}$, CVs serve both the demand $b^{rep}$ and $b^{nor}$ as many as possible, and AVs serve the residual demand. But, note that only when the number of CVs is enough to serve $b^{rep}$ (i.e., $N^P(r_1)\geq m_{th1}$), CVs start to serve the demand $b^{nor}$ as many as possible. 

If the number of CVs $N$ is too small to serve all the demand $b^{rep}$ (i.e., $N^P(r_1)<m_{th1})$), we have the same $\bm{x}^{A\star}$ and $\bm{x}^{C\star}$ as in old case III. 

If the number of CVs $N$ is enough to serve all the demand $b^{rep}$ (i.e., $N^P(r_1)\geq m_{th1}$), then the additional CVs will be assigned to serve the demand $b^{nor}$. Note that we need $m_{th2}:=b_{11}\tau_{11}+b_{12}\tau_{12}+b_{21}\tau_{21}+b_{22}\tau_{22}+(b_{21}-b_{12})\tau_{12}+(b_{11}+b_{12})w^*$ to serve all the demand $b^{rep}$ and $b^{nor}$, and thus we have
\begin{subequations}
\begin{align}
\bm{x}^{C\star}\hspace{-3pt}=\hspace{-2pt}&\min(1,\frac{N^P(r_1)}{m_{th1}}) \left[\begin{array}{cc}
   0 & b_{21}-b_{12} \\
   0 & b_{22}-b_{12}\frac{q_{22}}{q_{21}}
    \end{array}
    \right]\nonumber\\ 
  &  \hspace{-3pt}+\hspace{-3pt}\max(0,\hspace{-2pt}\min(1,\hspace{-2pt}\frac{N^P\hspace{-2pt}(r_1\hspace{-2pt})\hspace{-3pt}-\hspace{-3pt}m_{th1}}{m_{th2}\hspace{-3pt}-\hspace{-3pt}m_{th1}}\hspace{-3pt})\hspace{-1pt}) \hspace{-3pt}\left[\begin{array}{cc}
   b_{11}\hspace{-3pt}+\hspace{-3pt}b_{12}& 0\\
   0 &b_{12}\hspace{-3pt}+\hspace{-3pt}b_{12}\frac{q_{22}}{q_{21}}
    \end{array}\hspace{-3pt}
    \right]\hspace{-3pt}, \nonumber
\end{align}
\end{subequations}
and \[\bm{x}^{A\star}=\left[\begin{array}{cc}
   b_{11}+b_{12}& b_{21}-b_{12}\\
   0 &b_{12}+b_{22}
    \end{array}
    \right]-\bm{x}^C.\]

\section{Numerical Results for $2\times 2$ Grid} \label{app:2grid}

\begin{table*}[htbp]\label{tab:allinitial}
\TABLE
{Comparing the obtained platform profits of our proposed algorithms for different initial points in the $10$ simulations with distinct randomly generated demand matrix $\bm{b}$.}
{
\begin{tabular}{cccccccccccccc} 
\toprule
Demand $\bm{b}$& 
\multicolumn{7}{c}{$\begin{pmatrix}
0 & 2 & 0 &1 \\
0 & 0 & 2 &1 \\
2 & 0 & 0 &1 \\
1 & 0 & 2 &0
\end{pmatrix}$}\\
\midrule
Initial point 
& AV-frist & $\bm{0}$ & $\frac{1}{4}\bm{b}$ & $\frac{1}{2}\bm{b}$ & $\frac{3}{4}\bm{b}$ & $\bm{b}$ & Std\\


Gradient-descent 
& 12.31& 11.76& 11.76 &11.93 & 11.76 &11.42 & 0.19\\

Bundle method 
& 12.31& 12.48&12.50 &12.31 &12.12 &12.48 & 0.16\\

Genetic 
& \multicolumn{6}{c}{12.48} & -\\

\midrule
\midrule
Demand $\bm{b}$& 
\multicolumn{7}{c}{$\begin{pmatrix}
0 & 1 & 1 &1 \\
2 & 0 & 1 &1 \\
1 & 0 & 0 &2 \\
0 & 0 & 2 &0
\end{pmatrix}$}\\
\midrule
Initial point 
& AV-frist & $\bm{0}$ & $\frac{1}{4}\bm{b}$ & $\frac{1}{2}\bm{b}$ & $\frac{3}{4}\bm{b}$ & $\bm{b}$ & Std \\


Gradient-descent 
& 10.89& 9.93& 10.26& 10.16& 10.51&9.39 & 0.42\\

Bundle method 
& 11.04 & 10.89& 11.02& 10.89& 11.02&10.89 & 0.07\\

Genetic 
& \multicolumn{6}{c}{10.99} & -\\
\midrule
\midrule
Demand $\bm{b}$& 
\multicolumn{7}{c}{$\begin{pmatrix}
0 & 0 & 2 &1 \\
0 & 0 & 2 &0 \\
2 & 0 & 0 &2 \\
2 & 0 & 1 &0
\end{pmatrix}$}\\
\midrule
Initial point 
& AV-frist & $\bm{0}$ & $\frac{1}{4}\bm{b}$ & $\frac{1}{2}\bm{b}$ & $\frac{3}{4}\bm{b}$ & $\bm{b}$ & Std\\


Gradient-descent 
& 13.19& 13.03& 12.97& 12.04& 13.18& 12.99 & 0.46\\

Bundle method 
& 12.80& 13.03 &12.80 &13.20 & 13.01&13.03 & 0.14\\

Genetic 
& \multicolumn{6}{c}{12.80} & -\\

\midrule
\midrule
Demand $\bm{b}$& 
\multicolumn{7}{c}{$\begin{pmatrix}
0 & 0 & 0 &1 \\
2 & 0 & 2 &1 \\
1 & 2 & 0 &2 \\
0 & 1 & 2 &0
\end{pmatrix}$}\\
\midrule
Initial point 
& AV-frist & $\bm{0}$ & $\frac{1}{4}\bm{b}$ & $\frac{1}{2}\bm{b}$ & $\frac{3}{4}\bm{b}$ & $\bm{b}$ & Std\\


Gradient-descent 
& 14.54& 10.72& 10.77&13.44&13.79&13.48 & 1.56\\

Bundle method 
& 14.57& 14.57& 14.53& 14.57& 14.38& 14.57 & 0.08\\

Genetic 
& \multicolumn{6}{c}{14.60} & -\\

\midrule
\midrule
Demand $\bm{b}$& 
\multicolumn{7}{c}{$\begin{pmatrix} 
0 & 0 & 2 &0 \\
2 & 0 & 2 &1 \\
1 & 0 & 0 &0 \\
0 & 0 & 2 &0
\end{pmatrix}$}\\
\midrule
Initial point 
& AV-frist & $\bm{0}$ & $\frac{1}{4}\bm{b}$ & $\frac{1}{2}\bm{b}$ & $\frac{3}{4}\bm{b}$ & $\bm{b}$ & Std\\


Gradient-descent 
& 9.10 & 9.10& 9.05 &9.09 &9.10 &9.09 &  0.019\\

Bundle method 
& 9.10&  9.10& 9.10 &9.10  &9.10  & 9.10 &9.10 & 0  \\

Genetic 
& \multicolumn{6}{c}{9.10} & -\\

\midrule
\midrule
Demand $\bm{b}$& 
\multicolumn{7}{c}{$\begin{pmatrix} 
0 & 2 & 1 &2 \\
0 & 0 & 1 &2 \\
1 & 2 & 0 &2 \\
0 & 2 & 2 &0
\end{pmatrix}$}\\
\midrule
Initial point 
& AV-frist & $\bm{0}$ & $\frac{1}{4}\bm{b}$ & $\frac{1}{2}\bm{b}$ & $\frac{3}{4}\bm{b}$ & $\bm{b}$ & Std\\


Gradient-descent 
& 13.89& 10.55 &12.70  & 13.52  & 12.60 & 10.72 & 1.31 \\

Bundle method 
& 14.08& 14.07 & 14.19 & 14.00 & 14.32 & 14.07 & 0.13 \\

Genetic 
& \multicolumn{6}{c}{14.77} & -\\
\bottomrule

\end{tabular}}
{}
\end{table*}

\begin{table*}[htbp]
\TABLE
{Comparing the obtained platform profits of our proposed algorithms for different initial points in the $10$ simulations with distinct randomly generated demand matrix $\bm{b}$.}
{
\begin{tabular}{cccccccccccccc} 
\toprule
Demand $\bm{b}$& 
\multicolumn{7}{c}{$\begin{pmatrix} 
0 & 1 & 2 &0 \\
0 & 0 & 1 &1 \\
0 & 0 & 0 &1 \\
2 & 0 & 0 &0
\end{pmatrix}$}\\
\midrule
Initial point 
& AV-frist & $\bm{0}$ & $\frac{1}{4}\bm{b}$ & $\frac{1}{2}\bm{b}$ & $\frac{3}{4}\bm{b}$ & $\bm{b}$ & Std\\


Gradient-descent 
& 9.14& 8.12 &9.10  & 8.40 & 8.74 & 8.42 & 0.38 \\

Bundle method 
& 9.14& 8.70 & 9.13 & 8.98 & 8.70 & 8.70 & 0.20 \\

Genetic 
& \multicolumn{6}{c}{9.14} & -\\

\midrule
\midrule
Demand $\bm{b}$& 
\multicolumn{7}{c}{$\begin{pmatrix} 
0 & 0 & 0 &0 \\
2 & 0 & 0 &1 \\
2 & 1 & 0 &0 \\
2 & 0 & 0 &0
\end{pmatrix}$}\\
\midrule
Initial point 
& AV-frist & $\bm{0}$ & $\frac{1}{4}\bm{b}$ & $\frac{1}{2}\bm{b}$ & $\frac{3}{4}\bm{b}$ & $\bm{b}$ & Std\\


Gradient-descent 
& 8.47 & 8.25 & 8.01 & 8.40 & 8.23 & 8.42 & 0.16 \\

Bundle method 
& 8.47& 8.43 & 8.45 & 8.44 & 8.28 & 8.43 & 0.07 \\

Genetic 
& \multicolumn{6}{c}{8.47} & -\\

\midrule
\midrule
Demand $\bm{b}$& 
\multicolumn{7}{c}{$\begin{pmatrix} 
0 & 1 & 1 &1 \\
2 & 0 & 1 &1 \\
2 & 1 & 0 &0 \\
0 & 2 & 1 &0
\end{pmatrix}$}\\
\midrule
Initial point 
& AV-frist & $\bm{0}$ & $\frac{1}{4}\bm{b}$ & $\frac{1}{2}\bm{b}$ & $\frac{3}{4}\bm{b}$ & $\bm{b}$ & Std\\


Gradient-descent 
& 12.10& 11.20 & 11.97 & 10.12 & 11.44 & 10.37 & 0.77 \\

Bundle method 
& 12.40& 12.40 & 12.37 & 12.50 & 12.40 & 12.40 & 0.05 \\

Genetic 
& \multicolumn{6}{c}{12.40} & -\\

\midrule
\midrule
Demand $\bm{b}$& 
\multicolumn{7}{c}{$\begin{pmatrix} 
0 & 0 & 2 &0 \\
2 & 0 & 0 &2 \\
1 & 0 & 0 &0 \\
2 & 0 & 1 &0
\end{pmatrix}$}\\
\midrule
Initial point 
& AV-frist & $\bm{0}$ & $\frac{1}{4}\bm{b}$ & $\frac{1}{2}\bm{b}$ & $\frac{3}{4}\bm{b}$ & $\bm{b}$ & Std\\


Gradient-descent 
& 9.44& 9.07 & 8.16 & 8.72 & 9.14 & 9.34 & 0.46 \\

Bundle method 
& 9.44& 9.40 & 9.19 & 8.70 & 9.27 & 9.40 & 0.29 \\

Genetic 
& \multicolumn{6}{c}{9.44} & -\\

\bottomrule

\end{tabular}}
{}
\end{table*}


\ACKNOWLEDGMENT{%
}

%
%
%



\end{APPENDICES}
\end{document}